\documentclass[11pt]{article}
\usepackage{axodraw}
\usepackage{epsfig}
\usepackage{amsfonts}
\usepackage{amsmath}
\usepackage{bbm,bm}
\usepackage{cite}
\allowdisplaybreaks[1]

\input pix.sty

\renewcommand{\vec}[1]{{\bf #1}}

\newcommand{\alphas}{\alpha_{\rm s}}
\newcommand{\Nf}{N_{\rm f}}
\newcommand{\Nc}{N_{\rm c}}

\newcommand{\Tc}{T_{\rm c}}

\newcommand{\bg}{\rmii{$B$}}

\newcommand{\gB}{g_\rmii{B}}
\newcommand{\mE}{m_\rmii{E}}
\newcommand{\mER}{m_\rmii{ER}}

\newcommand{\mG}{m_\rmii{G}}
\newcommand{\gE}{g_\rmii{E}}
\newcommand{\gER}{g_\rmii{ER}}
\newcommand{\lE}{\lambda_\rmii{E}}
\newcommand{\kE}{\kappa_\rmii{E}}
\newcommand{\mEs}{m_\rmiii{E}}
\newcommand{\gEs}{g_\rmiii{E}}

\newcommand{\gM}{g_\rmii{M}}

\newcommand{\gammaE}{\gamma_\rmii{E}}
\newcommand{\varzeta}{\eta}
\newcommand{\rmO}{{\mathcal{O}}}
\newcommand{\bmu}{\bar\mu}

\def\lsi{\raise0.3ex\hbox{$<$\kern-0.75em\raise-1.1ex\hbox{$\sim$}}}
\def\gsi{\raise0.3ex\hbox{$>$\kern-0.75em\raise-1.1ex\hbox{$\sim$}}}
\newcommand{\lsim}{\mathop{\lsi}}

\newcommand{\rmii}[1]{{\mbox{\tiny\rm{#1}}}}
\newcommand{\rmiii}[1]{{\mbox{\tiny{$\scriptstyle{\rm#1}$}}}}

\newcommand{\Tint}[1]{{\hbox{$\sum$}\!\!\!\!\!\!\!\int\,}_{\!\!\!\!\raise-0.9ex\hbox{$\scriptstyle{#1}$}}}
\newcommand{\Tinti}[1]{{{\Sigma}\!\!\!\!\raise0.3ex\hbox{$\int$}_\rmii{${#1}$}}}
\newcommand{\Tintip}[1]{{{\Sigma'}\!\!\!\!\!\raise0.3ex\hbox{$\int$}_\rmii{${#1}$}}}

\newcommand{\bi}{\begin{itemize}}
\newcommand{\ei}{\end{itemize}}
\newcommand{\hide}[1]{ }

\def\TAsc(#1,#2)(#3,#4,#5)%
{\SetWidth{2.0}\CArc(#1,#2)(#3,#4,#5)\SetWidth{1.0}}
\def\Lwidth{3}

\def\TAgl(#1,#2)(#3,#4,#5){\SetWidth{2.0}\PhotonArc(#1,#2)(#3,#4,#5){\Lwidth}%
{6.283 #3 mul 360 div #4 #5 sub #4 #5 sub mul sqrt mul Tdensity mul}%
\SetWidth{1.0}}
\def\TLgl(#1,#2)(#3,#4){\SetWidth{2.0}\Photon(#1,#2)(#3,#4){\Lwidth}
{#1 #3 sub #1 #3 sub mul #2 #4 sub #2 #4 sub mul add sqrt Tdensity mul}%
\SetWidth{1.0}}
\def\Aegl(#1,#2)(#3,#4,#5){\PhotonArc(#1,#2)(#3,#4,#5){\Lwidth}
{6.283 #3 mul 360 div #4 #5 sub #4 #5 sub mul sqrt mul Ldensity mul}}
\def\Legl(#1,#2)(#3,#4){\Photon(#1,#2)(#3,#4){\Lwidth}
{#1 #3 sub #1 #3 sub mul #2 #4 sub #2 #4 sub mul add sqrt Ldensity mul}}
\def\ToprSBB(#1,#2,#3,#4,#5){\picb{#1(0,15)(7.5,15)  #1(37.5,15)(45,15)%
 #2(22.5,15)(15,0,70) #2(22.5,15)(15,110,180) #3(22.5,15)(15,180,360)%
 #4(22.5,30)(5,-10,190) #5(22.5,30)(5,190,350)}}
\def\ToprSBT(#1,#2,#3,#4){\picb{#1(0,15)(7.5,15)  #1(37.5,15)(45,15)%
 #2(22.5,15)(15,0,90) #2(22.5,15)(15,90,180) #3(22.5,15)(15,180,360)%
 #4(22.5,35)(5,-90,270)}}
\def\ToprSTB(#1,#2,#3,#4){\picb{#1(0,0)(22.5,0) #1(22.5,0)(45,0)%
 #2(22.5,15)(15,-90,70) #2(22.5,15)(15,110,270)%
 #3(22.5,30)(5,-10,190) #4(22.5,30)(5,190,350)}}
\def\ToprSTT(#1,#2,#3){\picb{#1(0,0)(22.5,0) #1(22.5,0)(45,0)%
 #2(22.5,15)(15,-90,90) #2(22.5,15)(15,90,270)%
 #3(22.5,35)(5,-90,270)}}
\def\ToptSMx(#1,#2,#3,#4,#5,#6){\picb{#1(0,15)(7.5,15) #1(37.5,15)(45,15)%
 #2(22.5,15)(15,0,90) #3(22.5,15)(15,90,180) #4(22.5,15)(15,180,270)%
 #5(22.5,15)(15,270,360) #6(22.5,30)(22.5,0)%
 \GCirc(11.9,25.6){2.5}{0}}}
\def\ToptSMy(#1,#2,#3,#4,#5,#6){\picb{#1(0,15)(7.5,15) #1(37.5,15)(45,15)%
 #2(22.5,15)(15,0,90) #3(22.5,15)(15,90,180) #4(22.5,15)(15,180,270)%
 #5(22.5,15)(15,270,360) #6(22.5,30)(22.5,0)%
 \GCirc(22.5,15){2.5}{0}}}
\def\ToptSMz(#1,#2,#3,#4,#5,#6){\picb{#1(0,15)(7.5,15) #1(37.5,15)(45,15)%
 #2(22.5,15)(15,0,90) #3(22.5,15)(15,90,180) #4(22.5,15)(15,180,270)%
 #5(22.5,15)(15,270,360) #6(22.5,30)(22.5,0)%
 \GCirc(33.1,25.6){2.5}{0}}}
\def\ToptSMa(#1,#2,#3,#4,#5,#6){\picb{#1(0,15)(7.5,15) #1(37.5,15)(45,15)%
 #2(22.5,15)(15,0,90) #3(22.5,15)(15,90,180) #4(22.5,15)(15,180,270)%
 #5(22.5,15)(15,270,360) #6(22.5,30)(22.5,0)%
 \GCirc(7.5,15){2.5}{0}}}
\def\ToptSMb(#1,#2,#3,#4,#5,#6){\picb{#1(0,15)(7.5,15) #1(37.5,15)(45,15)%
 #2(22.5,15)(15,0,90) #3(22.5,15)(15,90,180) #4(22.5,15)(15,180,270)%
 #5(22.5,15)(15,270,360) #6(22.5,30)(22.5,0)%
 \GCirc(22.5,30){2.5}{0}}}
\def\ToprSBBx(#1,#2,#3,#4,#5){\picb{#1(0,15)(7.5,15)  #1(37.5,15)(45,15)%
 #2(22.5,15)(15,0,70) #2(22.5,15)(15,110,180) #3(22.5,15)(15,180,360)%
 #4(22.5,30)(5,-10,190) #5(22.5,30)(5,190,350)%
 \GCirc(11.9,25.6){2.5}{0}}}
\def\ToprSBBy(#1,#2,#3,#4,#5){\picb{#1(0,15)(7.5,15)  #1(37.5,15)(45,15)%
 #2(22.5,15)(15,0,70) #2(22.5,15)(15,110,180) #3(22.5,15)(15,180,360)%
 #4(22.5,30)(5,-10,190) #5(22.5,30)(5,190,350)%
 \GCirc(22.5,0){2.5}{0}}}
\def\ToprSBBz(#1,#2,#3,#4,#5){\picb{#1(0,15)(7.5,15)  #1(37.5,15)(45,15)%
 #2(22.5,15)(15,0,70) #2(22.5,15)(15,110,180) #3(22.5,15)(15,180,360)%
 #4(22.5,30)(5,-10,190) #5(22.5,30)(5,190,350)%
 \GCirc(22.5,25){2.5}{0}}}
\def\ToprSBBw(#1,#2,#3,#4,#5){\picb{#1(0,15)(7.5,15)  #1(37.5,15)(45,15)%
 #2(22.5,15)(15,0,70) #2(22.5,15)(15,110,180) #3(22.5,15)(15,180,360)%
 #4(22.5,30)(5,-10,190) #5(22.5,30)(5,190,350)%
 \GCirc(22.5,35){2.5}{0}}}
\def\ToprSBBa(#1,#2,#3,#4,#5){\picb{#1(0,15)(7.5,15)  #1(37.5,15)(45,15)%
 #2(22.5,15)(15,0,70) #2(22.5,15)(15,110,180) #3(22.5,15)(15,180,360)%
 #4(22.5,30)(5,-10,190) #5(22.5,30)(5,190,350)%
 \GCirc(7.5,15){2.5}{0}}}
\def\ToprSBBb(#1,#2,#3,#4,#5){\picb{#1(0,15)(7.5,15)  #1(37.5,15)(45,15)%
 #2(22.5,15)(15,0,70) #2(22.5,15)(15,110,180) #3(22.5,15)(15,180,360)%
 #4(22.5,30)(5,-10,190) #5(22.5,30)(5,190,350)%
 \GCirc(17.5,29){2.5}{0}}}
\def\ToptSAlx(#1,#2,#3,#4,#5){\picb{#1(0,15)(7.5,15) #1(37.5,15)(45,15)%
 #2(22.5,15)(15,0,90) #3(22.5,15)(15,90,180) #4(22.5,15)(15,180,360)%
 #5(7.5,30)(15,270,360)%
 \GCirc(11.9,25.6){2.5}{0}}}
\def\ToptSAlz(#1,#2,#3,#4,#5){\picb{#1(0,15)(7.5,15) #1(37.5,15)(45,15)%
 #2(22.5,15)(15,0,90) #3(22.5,15)(15,90,180) #4(22.5,15)(15,180,360)%
 #5(7.5,30)(15,270,360)%
 \GCirc(33.1,25.6){2.5}{0}}}
\def\ToptSAly(#1,#2,#3,#4,#5){\picb{#1(0,15)(7.5,15) #1(37.5,15)(45,15)%
 #2(22.5,15)(15,0,90) #3(22.5,15)(15,90,180) #4(22.5,15)(15,180,360)%
 #5(7.5,30)(15,270,360)%
 \GCirc(22.5,0){2.5}{0}}}
\def\ToptSAlw(#1,#2,#3,#4,#5){\picb{#1(0,15)(7.5,15) #1(37.5,15)(45,15)%
 #2(22.5,15)(15,0,90) #3(22.5,15)(15,90,180) #4(22.5,15)(15,180,360)%
 #5(7.5,30)(15,270,360)%
 \GCirc(17.2,20.3){2.5}{0}}}
\def\ToptSAla(#1,#2,#3,#4,#5){\picb{#1(0,15)(7.5,15) #1(37.5,15)(45,15)%
 #2(22.5,15)(15,0,90) #3(22.5,15)(15,90,180) #4(22.5,15)(15,180,360)%
 #5(7.5,30)(15,270,360)%
 \GCirc(22.5,30){2.5}{0}}}
\def\ToptSAlb(#1,#2,#3,#4,#5){\picb{#1(0,15)(7.5,15) #1(37.5,15)(45,15)%
 #2(22.5,15)(15,0,90) #3(22.5,15)(15,90,180) #4(22.5,15)(15,180,360)%
 #5(7.5,30)(15,270,360)%
 \GCirc(37.5,15){2.5}{0}}}
\def\ToptSAlc(#1,#2,#3,#4,#5){\picb{#1(0,15)(7.5,15) #1(37.5,15)(45,15)%
 #2(22.5,15)(15,0,90) #3(22.5,15)(15,90,180) #4(22.5,15)(15,180,360)%
 #5(7.5,30)(15,270,360)%
 \GCirc(7.5,15){2.5}{0}}}
\def\ToprSTBx(#1,#2,#3,#4){\picb{#1(0,0)(22.5,0) #1(22.5,0)(45,0)%
 #2(22.5,15)(15,-90,70) #2(22.5,15)(15,110,270)%
 #3(22.5,30)(5,-10,190) #4(22.5,30)(5,190,350)%
 \GCirc(7.5,15){2.5}{0}}}
\def\ToprSTBy(#1,#2,#3,#4){\picb{#1(0,0)(22.5,0) #1(22.5,0)(45,0)%
 #2(22.5,15)(15,-90,70) #2(22.5,15)(15,110,270)%
 #3(22.5,30)(5,-10,190) #4(22.5,30)(5,190,350)%
 \GCirc(22.5,25){2.5}{0}}}
\def\ToprSTBz(#1,#2,#3,#4){\picb{#1(0,0)(22.5,0) #1(22.5,0)(45,0)%
 #2(22.5,15)(15,-90,70) #2(22.5,15)(15,110,270)%
 #3(22.5,30)(5,-10,190) #4(22.5,30)(5,190,350)%
 \GCirc(22.5,35){2.5}{0}}}
\def\ToprSTBa(#1,#2,#3,#4){\picb{#1(0,0)(22.5,0) #1(22.5,0)(45,0)%
 #2(22.5,15)(15,-90,70) #2(22.5,15)(15,110,270)%
 #3(22.5,30)(5,-10,190) #4(22.5,30)(5,190,350)%
 \GCirc(17.5,29){2.5}{0}}}
\def\ToprSTBc(#1,#2,#3,#4){\picb{#1(0,0)(22.5,0) #1(22.5,0)(45,0)%
 #2(22.5,15)(15,-90,70) #2(22.5,15)(15,110,270)%
 #3(22.5,30)(5,-10,190) #4(22.5,30)(5,190,350)%
 \GCirc(22.5,0){2.5}{0}}}
\def\ToptSSx(#1,#2,#3,#4){\picb{#1(0,15)(7.5,15) #1(37.5,15)(45,15)%
 #4(7.5,15)(37.5,15) #2(22.5,15)(15,0,180) #3(22.5,15)(15,180,360)%
 \GCirc(22.5,15){2.5}{0}}}
\def\ToptSSy(#1,#2,#3,#4){\picb{#1(0,15)(7.5,15) #1(37.5,15)(45,15)%
 #4(7.5,15)(37.5,15) #2(22.5,15)(15,0,180) #3(22.5,15)(15,180,360)%
 \GCirc(22.5,30){2.5}{0}}}
\def\ToptSSc(#1,#2,#3,#4){\picb{#1(0,15)(7.5,15) #1(37.5,15)(45,15)%
 #4(7.5,15)(37.5,15) #2(22.5,15)(15,0,180) #3(22.5,15)(15,180,360)%
 \GCirc(7.5,15){2.5}{0}}}
\def\ToprSBTc(#1,#2,#3,#4){\picb{#1(0,15)(7.5,15)  #1(37.5,15)(45,15)%
 #2(22.5,15)(15,0,90) #2(22.5,15)(15,90,180) #3(22.5,15)(15,180,360)%
 #4(22.5,35)(5,-90,270)%
 \GCirc(22.5,30){2.5}{0}}}
\def\ToprSTTc(#1,#2,#3){\picb{#1(0,0)(22.5,0) #1(22.5,0)(45,0)%
 #2(22.5,15)(15,-90,90) #2(22.5,15)(15,90,270)%
 #3(22.5,35)(5,-90,270)%
 \GCirc(22.5,30){2.5}{0}}}
\def\ToptSEc(#1,#2,#3,#4,#5){\picb{#1(0,15)(7.5,15) #1(37.5,15)(45,15)%
 #3(15,15)(7.5,0,180) #4(15,15)(7.5,180,360)%
 #2(30,15)(7.5,0,180) #5(30,15)(7.5,180,360)%
 \GCirc(22.5,15){2.5}{0}}} 

\renewcommand{\picc}[1]{\;\parbox[c]{60pt}{\begin{picture}(60,30)(0,0)
\SetWidth{1.0}\SetScale{1.3} #1 \end{picture}}\; }
\def\Lwidth{1.3}

\newcommand{\picu}[1]{\;\parbox[c]{40pt}{\begin{picture}(50,30)(0,0)
\SetWidth{1.0}\SetScale{0.5} #1 \end{picture}}\; }

\def\ScatA{\picu{%
 \SetWidth{2.0} 
 \Photon(0,30)(30,30){2}{3}%
 \Photon(60,30)(30,30){-2}{3}%
 \GCirc(30,60){3}{0}%
 \CArc(30,45)(15,0,360)%
}}
\def\ScatB{\picu{%
 \SetWidth{2.0} 
 \Photon(-15,30)(15,30){2}{3}%
 \Photon(75,30)(45,30){-2}{3}%
 \GCirc(30,45){3}{0}%
 \CArc(30,30)(15,0,360)%
}}
\def\ScatC{\picu{%
 \SetWidth{2.0} 
 \Photon(0,30)(30,30){2}{3}%
 \Photon(60,30)(30,30){-2}{3}%
 \GCirc(30,30){3}{0}%
 \CArc(30,45)(15,0,360)%
}}
\def\ScatD{\picu{%
 \SetWidth{2.0} 
 \Photon(-15,30)(15,30){2}{3}%
 \Photon(75,30)(45,30){-2}{3}%
 \GCirc(15,30){3}{0}%
 \CArc(30,30)(15,0,360)%
}}
\def\DiagA{\picu{%
 \SetWidth{2.0} 
 \PhotonArc(30,30)(15,0,360){1.5}{15}%
 \Photon(45,30)(60,30){1.5}{3}%
 \Photon(34.64,44.27)(39.27,58.53){1.5}{3}%
 \Photon(17.86,38.82)(5.73,47.63){1.5}{3}%
 \Photon(17.86,21.18)(5.73,12.37){1.5}{3}%
 \Photon(34.64,15.73)(39.27,1.47){1.5}{3}%
}}
\def\DiagB{\picu{%
 \SetWidth{2.0} 
 \DashArrowArc(30,30)(15,0,360){2}%
 \Photon(45,30)(60,30){1.5}{3}%
 \Photon(34.64,44.27)(39.27,58.53){1.5}{3}%
 \Photon(17.86,38.82)(5.73,47.63){1.5}{3}%
 \Photon(17.86,21.18)(5.73,12.37){1.5}{3}%
 \Photon(34.64,15.73)(39.27,1.47){1.5}{3}%
}}
\def\DiagBsc{\picu{%
 \SetWidth{2.0} 
 \CArc(30,30)(15,0,360)%
 \Photon(45,30)(60,30){1.5}{3}%
 \Photon(34.64,44.27)(39.27,58.53){1.5}{3}%
 \Photon(17.86,38.82)(5.73,47.63){1.5}{3}%
 \Photon(17.86,21.18)(5.73,12.37){1.5}{3}%
 \Photon(34.64,15.73)(39.27,1.47){1.5}{3}%
}}
\def\DiagBma{\picu{%
 \SetWidth{2.0} 
 \CArc(30,30)(15,0,360)%
 \Photon(45,30)(60,30){1.5}{3}%
 \Photon(15,30)(0,30){1.5}{3}%
}}
\def\DiagBmb{\picu{%
 \SetWidth{2.0} 
 \CArc(30,45)(15,0,360)%
 \Photon(30,30)(55,30){1.5}{5}%
 \Photon(30,30)(5,30){1.5}{5}%
}}
\def\DiagBmc{\picu{%
 \SetWidth{2.0} 
 \CArc(30,30)(15,0,360)%
 \Photon(45,30)(60,30){1.5}{3}%
 \Photon(22.5,42.99)(15,55.98){1.5}{3}%
 \Photon(22.5,17.01)(15,4.02){1.5}{3}%
}}
\def\DiagC{\picu{%
 \SetWidth{2.0} 
 \PhotonArc(30,30)(15,0,360){1.5}{15}%
 \Photon(45,30)(57.12,38.82){1.5}{3}%
 \Photon(45,30)(57.12,21.18){-1.5}{3}%
 \Photon(30,45)(30,60){1.5}{3}%
 \Photon(14,30)(-1,30){1.5}{3}%
 \Photon(30,15)(30,0){1.5}{3}%
}}
\def\DiagD{\picu{%
 \SetWidth{2.0} 
 \DashArrowArc(30,30)(15,-45,315){2}%
 \Photon(45,30)(57.12,38.82){1.5}{3}%
 \Photon(45,30)(57.12,21.18){-1.5}{3}%
 \Photon(30,45)(30,60){1.5}{3}%
 \Photon(15,30)(0,30){1.5}{3}%
 \Photon(30,15)(30,0){1.5}{3}%
}}
\def\DiagDsc{\picu{%
 \SetWidth{2.0} 
 \CArc(30,30)(15,-45,315)%
 \Photon(45,30)(57.12,38.82){1.5}{3}%
 \Photon(45,30)(57.12,21.18){-1.5}{3}%
 \Photon(30,45)(30,60){1.5}{3}%
 \Photon(15,30)(0,30){1.5}{3}%
 \Photon(30,15)(30,0){1.5}{3}%
}}
\def\DiagBmd{\picu{%
 \SetWidth{2.0} 
 \CArc(30,30)(15,-45,315)%
 \Photon(15,30)(2.88,38.82){1.5}{3}%
 \Photon(15,30)(2.88,21.18){-1.5}{3}%
 \Photon(45,30)(60,30){1.5}{3}%
}}
\def\DiagE{\picu{%
 \SetWidth{2.0} 
 \PhotonArc(30,30)(15,0,360){1.5}{15}%
 \Photon(45,30)(60,30){1.5}{3}%
 \Photon(22.5,42.99)(26.38,57.48){1.5}{3}%
 \Photon(22.5,42.99)(8.01,46.87){-1.5}{3}%
 \Photon(22.5,17.01)(26.38,2.52){-1.5}{3}%
 \Photon(22.5,17.01)(8.01,13.13){1.5}{3}%
}}
\def\DiagF{\picu{%
 \SetWidth{2.0} 
 \DashArrowArc(30,30)(15,0,360){2}%
 \Photon(45,30)(60,30){1.5}{3}%
 \Photon(22.5,42.99)(26.38,57.48){1.5}{3}%
 \Photon(22.5,42.99)(8.01,46.87){-1.5}{3}%
 \Photon(22.5,17.01)(26.38,2.52){-1.5}{3}%
 \Photon(22.5,17.01)(8.01,13.13){1.5}{3}%
}}
\def\DiagFsc{\picu{%
 \SetWidth{2.0} 
 \CArc(30,30)(15,0,360)%
 \Photon(45,30)(60,30){1.5}{3}%
 \Photon(22.5,42.99)(26.38,57.48){1.5}{3}%
 \Photon(22.5,42.99)(8.01,46.87){-1.5}{3}%
 \Photon(22.5,17.01)(26.38,2.52){-1.5}{3}%
 \Photon(22.5,17.01)(8.01,13.13){1.5}{3}%
}}
\def\FiveA{\picu{%
 \SetWidth{2.0} 
 \Photon(0,30)(30,30){2}{3}%
 \Photon(60,30)(30,30){-2}{3}%
 \GCirc(30,30){3}{0}%
 \CArc(30,50)(20,0,360)%
 \Photon(30,70)(30,30){2}{5}%
}}
\def\FiveB{\picu{%
 \SetWidth{2.0} 
 \Photon(-15,30)(15,30){2}{3}%
 \Photon(85,30)(55,30){-2}{3}%
 \GCirc(15,30){3}{0}%
 \CArc(35,30)(20,0,360)%
 \CArc(26,30)(10,0,360)%
}}
\def\SixA{\picu{%
 \SetWidth{2.0} 
 \Photon(0,30)(30,30){2}{3}%
 \Photon(60,30)(30,30){-2}{3}%
 \GCirc(30,30){3}{0}%
 \CArc(30,43)(12,0,360)%
 \CArc(30,17)(12,0,360)%
}}
\makeatletter \@addtoreset{equation}{section} \makeatother
\renewcommand{\theequation}{\arabic{section}.\arabic{equation}}
\makeatletter
\renewcommand\section{\@startsection {section}{1}{\z@}%
                                   {-5.5ex \@plus -1ex \@minus -.2ex}
                                   {2.3ex \@plus.2ex}%
                                   {\normalfont\large\bfseries}}
\renewcommand\subsection{\@startsection{subsection}{2}{\z@}%
                                     {-3.25ex\@plus -1ex \@minus -.2ex}%
                                     {1.5ex \@plus .2ex}%
                                     {\normalfont\normalsize\bfseries}}
\renewcommand\thesection {\@arabic\c@section}
\renewcommand\thesubsection   {\thesection.\@arabic\c@subsection}
\renewcommand{\@seccntformat}[1]{%
\csname the#1\endcsname.\hspace{1.0em}}
\makeatother

\begin{document}

\flushbottom

\begin{titlepage}

\begin{flushright}
April 2018
\end{flushright}
\begin{centering}

\vfill

{\Large{\bf
 Soft thermal contributions to 3-loop gauge coupling 
}} 

\vspace{0.8cm}

M.~Laine$^{\rm a}$, 
P.~Schicho$^{\rm a}$,
Y.~Schr\"oder$^{\rm b}$

\vspace{0.8cm}

$^\rmi{a}$%
{\em
 AEC, Institute for Theoretical Physics, University of Bern, \\[1mm] 
 Sidlerstrasse 5, 3012 Bern, Switzerland\\
}

\vspace*{0.3cm}

$^\rmi{b}$%
{\em
 Grupo de Cosmolog\'ia y Part\'iculas Elementales,
 Universidad del B\'io-B\'io, \\[1mm]
 Casilla 447, Chill\'an, Chile
}

\vspace*{0.8cm}

\mbox{\bf Abstract}
 
\end{centering}

\vspace*{0.3cm}
 
\noindent
We analyze 3-loop contributions to the gauge coupling felt by ultrasoft 
(``magnetostatic'') modes in hot Yang-Mills theory. So-called soft/hard 
terms, originating from dimension-six operators within the soft effective 
theory, are shown to cancel 1097/1098 of the IR divergence found
in a recent determination of the hard 3-loop contribution to the soft 
gauge coupling. The remaining 1/1098 originates from ultrasoft/hard 
contributions, induced by dimension-six operators in the ultrasoft 
effective theory. Soft 3-loop contributions are likewise computed, 
and are found to be IR divergent, rendering the ultrasoft gauge 
coupling non-perturbative at relative order $\rmO(\alphas^{3/2})$. 
We elaborate on the implications of these findings for effective 
theory studies of physical observables in thermal QCD. 

\vfill

 
  
\vfill

\end{titlepage}

%
\section{Introduction} 

Dimensionally reduced (``3d'') thermal
effective theories, 
originally conceived for studying  
thermodynamics and phase transitions
in non-Abelian gauge theories~\cite{dr1,dr2,generic}, 
and still used for that purpose in the
context of weak interactions (cf.\ e.g.\ refs.~\cite{weir,weir2} for
recent work and references), have been reinvigorated in another
context some time ago. Indeed, quite remarkably, 
they also turn out to determine soft contributions 
to real-time lightcone observables~\cite{sch}. 
As examples, they can be used for estimating 
the so-called transverse collision kernel related 
to jet quenching in a hot QCD plasma~\cite{panero,dono}; 
soft parts of the photon and dilepton production rates from a 
QCD plasma~\cite{photon,dilepton}; 
and the interaction rate experienced by 
neutrinos in an electroweak plasma~\cite{broken}. 
Following standard terminology, we refer to the ``soft'' 
effective theory as EQCD, whereas the ``ultrasoft'' theory
containing only the magnetostatic modes is called MQCD 
(cf.\ e.g.\ refs.~\cite{linde,gpy,nadkarni,bn}). The latter
has been argued to give e.g.\ the leading non-perturbative
contribution to jet quenching~\cite{qhat}. 

In the QCD context 
it is known, however, that EQCD fails to
describe the full theory close to the phase transition or crossover 
temperature ($\Tc^{ }$). This is obvious when light quarks are present: 
EQCD contains only gluonic
degrees of freedom, and displays no remnant of 
the flavour symmetries that underlie
the chiral transition.  For  pure-glue theory,  
the reason for the breakdown is more subtle. 
Even though the center symmetry that drives the transition
in the imaginary-time formulation~\cite{sy} 
is not explicit in EQCD, remnants
of it are generated dynamically~\cite{adjoint}.
However the dynamical re-generation is incomplete, 
and a 3d lattice study in which soft EQCD dynamics 
was treated non-perturbatively did not achieve satisfactory agreement 
with thermodynamic functions obtained
from full 4d lattice simulations~\cite{lattg7}. 

One purpose of this paper is to demonstrate analytically 
that power-suppressed dimension-six operators, truncated 
from the super-renormalizable EQCD description, play an essential role 
in soft and ultrasoft observables, and are therefore a likely culprit
for EQCD's failure close to $\Tc^{ }$. More concretely, we determine
the MQCD gauge coupling in terms of the EQCD gauge coupling and mass
parameter up to 3-loop level, including the 1- and 2-loop 
contributions of all dimension-six operators; the result is 
contained in \eqs\nr{finalZ2}, \nr{finaldZ2} and \nr{gM2_3l}.  

Our presentation is organized as follows. 
After reviewing the form of EQCD and 
re-deriving the coefficients of its 
dimension-six operators 
in \se\ref{se:chapman}, we determine overlapping
soft/hard and ultrasoft/hard contributions to the  
ultrasoft gauge coupling in \se\ref{se:scalepiT}. 
In terms of four-dimensional Yang-Mills 
we go up to 3-loop level; this implies
2-loop level in effects originating from dimension-six operators, 
which are themselves generated by 1-loop diagrams. 
A 3-loop computation of soft effects, as well as of overlapping
ultrasoft/soft contributions, 
is presented in \se\ref{se:scalemE}, whereas 
conclusions are collected in \se\ref{se:concl}. 
Spacetime and colour tensors, tensor-like 1-loop sum-integrals, 
Feynman rules related to dimension-six operators, $d$-dimensional 
vacuum integrals, and some lengthier results, 
are collected in five appendices, respectively. 

%
\section{Form of EQCD}
\la{se:chapman}

%
\subsection{Super-renormalizable part}
\la{ss:SEQCD}

The super-renormalizable truncation of the dimensionally reduced
``electrostatic'' QCD, called EQCD, is defined by the action
\ba
 S^{ }_\rmii{EQCD}[A] & \equiv & 
 \int_X \biggl\{  
  \fr14 F^a_{ij}F^a_{ij}
 +
  \fr12 \mathcal{D}^{ab}_i\! A^b_0\, \mathcal{D}^{ac}_i\! A^c_0 
 + 
  \frac{\mE^2}{2} A_0^a A_0^a 
 \nn 
 & + &  
 \frac{\lE^{ }}{4} X^{abcd}_{ } A_0^a A_0^b A_0^c A_0^d
 + \frac{\kE^{ }}{4} A_0^a A_0^a A_0^b A_0^b
 \biggr\}
 \;. \la{S_EQCD}
\ea
Here $\int_X \equiv \frac{1}{T} \int_{\vec{x}}$, 
$
 F^a_{ij} \equiv \partial^{ }_i A^a_j - \partial^{ }_j A^a_i
 + \gE^{ } f^{abc} A^b_i A^c_j 
$,
$\mathcal{D}^{ab}_i \equiv \delta^{ab} \partial_i - \gE^{ } f^{abc}A^c_i$, 
$A_0^a$ is an adjoint scalar,
$X^{abcd}_{ }$ is defined in \eq\nr{def_X},  
Latin indices take values $i,j\in\{1,...,d\}$,
we have in mind $d \equiv 3 - 2\epsilon$, 
and repeated indices are summed over. We employ a convention in which
the fields $A^a_i$ and $A^a_0$ have the same dimensionality as
in four-dimensional Yang-Mills theory. 
Then explicit factors of $1/T$ and $T$ appear in configuration and momentum
space integration measures, respectively, 
where $T$ is the temperature.
  
Focussing on pure SU($\Nc^{ }$) gauge theory,\footnote{%
 We omit fermions for simplicity
 because they carry non-zero Matsubara freqencies and
 thus generate no direct IR divergences. 
 In other words they have no bearing
 on our conceptual discussion. If they were to be included,
 the expressions in \eqs\nr{mE}--\nr{lE}, \nr{c1}--\nr{c8}
 and, most importantly, \nr{cZbg_3loop}--\nr{ioan_eps},
 would contain additional terms involving $\Nf^{ }$. Unfortunately
 the determination of the last of these effects entails an enormous
 practical effort, which we defer to future work.   
 } 
i.e.\ suppressing contributions proportional to the number of
fermion flavours ($\Nf^{ }$), 
the parameters appearing in \eq\nr{S_EQCD}
have the expressions  
\ba
 \mE^2 & = & 
 \gB^2 \Nc^{ }\,  \Tint{P}'\frac{(d-1)^2}{P^2} 
 + \rmO\bigl(\gB^4\bigr) 
 \;, \la{mE} \\ 
 \gE^2 & = & 
 \gB^2 \,
 \biggl[  
  1 + \gB^2 \Nc^{ }\, \Tint{P}' \frac{25-d}{6P^4}
  + \rmO\bigl(\gB^4\bigr) 
 \biggr] 
 \;, \la{gE} \\ 
 \lE^{ } & = & 
 \gB^4 (d-1)^2 (3-d) \Tint{P}' \frac{1}{3 P^4}
 + \rmO\bigl(\gB^6\bigr) 
 \;, \la{lE} \quad
 \kE^{ } \; = \; \rmO\bigl( \gB^4\Nf^{ } \bigr) 
 \;, 
\ea
where $g^2_\rmii{B} = g^2 \mu^{2\epsilon}_{ } (1 + \rmO(g^2))$ 
is the bare coupling of the original four-dimensional theory, 
$\mu$ is the scale 
parameter introduced in the context of dimensional regularization, 
and $g^2 \equiv 4\pi\alphas$ is the renormalized
coupling. By $\Tintip{P}$ we denote 
a sum-integral over $P$, with the prime indicating that
the Matsubara zero mode is omitted. A 1-loop re-derivation of 
\eqs\nr{mE}--\nr{lE} can be found as a side product of
\se\ref{ss:determine}; 
2-loop expressions were obtained in ref.~\cite{gE2}; 
the 3-loop level has been reached
for $\mE^2$~\cite{ig_mE} and $\gE^2$~\cite{ig,ig_gE}.  

For our higher-loop computations in \se\ref{se:scalepiT}, 
it is helpful to express the dependence on $\lE^{ }$ and $\kE^{ }$
through the dimensionless combinations  
\ba
 \lambda & \equiv & 
 \frac{5 \lE^{ }\Nc^{ }}{4 \gE^2}
 + \frac{\kE^{ } ( \Nc^2+1) }{2\gE^2\Nc^{ }}    
 \;, \la{sc_a} \\ 
 \kappa^{ }_1 & \equiv & 
 \frac{\lE^{ } (\Nc^2 + 36 ) }{2 \gE^2 \Nc^{ }} 
 + \frac{10 \kE^{ }}{\gE^2\Nc^{ }}
 \;, \\ 
 \kappa^{ }_2 & \equiv & 
  \frac{\lE^2 ( \Nc^2 + 36 ) }{4 \gE^4} 
 +  \frac{10 \lE^{ } \kE^{ } }{\gE^4}
 + \frac{2 \kE^2(\Nc^2+1)}{\gE^4 \Nc^2}
 \;.  \la{sc_c}
\ea
We note in passing that  
fundamental representation couplings often used
in the literature, 
{\em viz.}
$
 \lE^{(1)} (\tr[A_0^2])^2 + \lE^{(2)} \tr[A_0^4]
$,  
are given by 
$\lE^{(1)} = 3 \lE^{ }/2 + \kE^{ }$ and $\lE^{(2)} = \lE^{ }\Nc^{ }/2$.

The theory can be renormalized through
\be
 \gE^2 \; = \; \gER^2 \mu^{2\epsilon} + \delta \gE^2
 \;, \quad 
 \mE^2 \; = \; m_\rmii{ER}^2 + \delta \mE^2
 \;, \la{gE_ct}
\ee
and similarly for the scalar couplings. 
Within the super-renormalizable truncation, 
the counterterms take the forms~\cite{pert,contlatt} 
\be
 \delta \gE^2 = 0
 \;, \quad
 \delta \mE^2 
 \; = \; 
 \biggl( \frac{\gER^2 \Nc^{ } T }{4\pi} \biggr)^2_{ }
 \frac{\kappa^{ }_2 - 4 \lambda}{4\epsilon}
 \;. \la{mE_ct}
\ee

The starting point for our analysis is the 3-loop determination 
of $\gE^2$ from four-dimensional Yang-Mills theory~\cite{ig,ig_gE}. 
It is helpful to display the result in the form 
of a background field effective action~\cite{lfa}.  
After gauge coupling and wave function renormalization through
vacuum counterterms, 
refs.~\cite{ig,ig_gE} found an expression containing
a logarithmic ($1/\epsilon$) divergence, 
\ba
 \Gamma^{(2)}_\rmii{EQCD}[B] 
  & = &   \frac{1}{2}\, B^a_i(q)\, B^b_j(r) 
 \, \delta^{ab} \, \delta(q+r) 
 \, \bigl( q^2 \delta^{ }_{ij} - q^{ }_i q^{ }_j\bigr)
 \, \bigl( \mathcal{Z}^{ }_{\bg}
  + \delta \mathcal{Z}^{ }_{\bg} \bigr) 
 \;, \la{calZ2} \\[2mm]
 \mathcal{Z}^{ }_{\bg} 
 & = & 
   1 - \frac{g^2 \Nc^{ }}{(4\pi)^2}
       \biggl[ \frac{22}{3} \ln \biggl(
           \frac{\bmu e^{\gammaE}}{4\pi T}
       \biggr) + \fr{1}{3} \biggr]
     - \frac{g^4 \Nc^2}{(4\pi)^4}
       \biggl[ \frac{68}{3} \ln \biggl(
           \frac{\bmu e^{\gammaE}}{4\pi T}
       \biggr) + \fr{341}{18} - \frac{10 \zeta^{ }_3}{9} \biggr]
 \la{cZbg_3loop} \\[2mm] 
 & - & 
   \frac{g^6\Nc^3}{(4\pi)^6} 
   \biggl[
   \frac{748}{9} \ln^2 \biggl(
           \frac{\bmu e^{\gammaE}}{4\pi T}
                       \biggr)
  +
  \biggl(  
    \frac{6608}{27} - \frac{10982 \zeta^{ }_3}{135}
  \biggr)
    \ln \biggl(
           \frac{\bmu e^{\gammaE}}{4\pi T}
                       \biggr)
  + \mbox{(finite)} \biggr]
 + \rmO(g^8) 
 \;, \nn[2mm]
 \delta \mathcal{Z}^{ }_{\bg} 
 & = & 
   \frac{g^6\Nc^3}{(4\pi)^6} 
   \frac{61\zeta^{ }_3}{5\epsilon}
 + \rmO(g^8) 
 \;. \la{ioan_eps}
\ea
Here $\zeta^{ }_n \equiv \zeta(n)$ and  
$\bmu^2 \equiv 4\pi \mu^2 e^{-\gammaE}$.
The renormalized gauge coupling is 
given by $\gER^2 = g^2/\mathcal{Z}^{ }_{\bg}$, 
and the corresponding counterterm by 
$\delta \gE^2 = - g^2 \mu^{2\epsilon} 
\delta \mathcal{Z}^{ }_{\bg} + \rmO(g^{10})$. 
We stress that \eqs\nr{cZbg_3loop} and \nr{ioan_eps}
are gauge independent~\cite{jm}.

An essential technical goal of our investigation is to demonstrate how the
divergence in \eq\nr{ioan_eps} 
is cancelled by overlapping soft/hard and ultrasoft/hard 
contributions, originating from dimension-six operators within 
EQCD and MQCD, respectively. 

At this point we would like to clarify why such logarithmic divergences
(which are ``universal'', i.e.\ present in any regularization scheme)
originate first at 3-loop level. In three dimensions, 1-loop graphs
may contain power divergences but no logarithmic divergences. Logarithmic
divergences first originate at 2-loop level. However, within the 
super-renormalizable truncation of EQCD, they lead to the counterterms
in \eq\nr{mE_ct}, i.e.\ the gauge coupling is finite. Divergences affecting
the gauge coupling can only emerge when dimension-six operators are added
to EQCD. 
Given that dimension-six operators are themselves generated by 1-loop
diagrams, the divergences correspond to the 3-loop level in terms of
the fundamental theory. In \se\ref{se:scalepiT}, 
where effects originating from integrating
out the hard scale are considered, 3-loop level corresponds to the relative
accuracy $O(g^6)$, whereas in \se\ref{se:scalemE}, 
where effects originating from
integrating out the soft scale are at focus, the expansion parameter is
$\sim g$, and the 3-loop effects are of relative magnitude $O(g^3)$.

%
\subsection{Dimension-six operators}

The dimension-six operators that can be added to \eq\nr{S_EQCD}
were determined in ref.~\cite{chapman}. 
We represent the operators as matrices
in the adjoint representation. Letting Greek indices
take values $\mu\in\{0,...,d\}$, computing the
coefficients at 1-loop level, and choosing to rephrase
the gauge coupling as the same $\gE^{ }$ as appears
inside $F^{a}_{ij}$ and $\mathcal{D}^{ab}_i$, the dimension-six action 
can be written as
\ba
 \delta S^{ }_\rmii{EQCD}[A] & = & 
 \Tint{P}' \frac{2 \gE^2 }{P^6} \int_X \mathop{\rm tr}
 \Bigl\{ 
  c^{ }_1 \, (D^{ }_\mu F^{ }_{\mu\nu})^2 + 
  c^{ }_2 \, (D^{ }_\mu F^{ }_{\mu 0})^2 
 \nn 
 & + & i\gE 
 \bigl[
  c^{ }_3\, F^{ }_{\mu\nu} F^{ }_{\nu\rho} F^{ }_{\rho\mu } + 
  c^{ }_4\, F^{ }_{0 \mu} F^{ }_{\mu\nu} F^{ }_{\nu 0  } + 
  c^{ }_5\, A^{ }_0 (D^{ }_{\mu} F^{ }_{\mu\nu}) F^{ }_{ 0 \nu}  
 \bigr]
 \nn[2mm] 
 & + & 
 \gE^2 
 \bigl[ 
  c^{ }_6\, A_0^2 F_{\mu\nu}^2 + 
  c^{ }_7 \, A^{ }_0 F^{ }_{\mu\nu} A^{ }_0 F^{ }_{\mu\nu} + 
  c^{ }_8 \, A_0^2 F_{0\mu}^2 +
  c^{ }_9 \, A^{ }_0 F^{ }_{0 \mu} A^{ }_0 F^{ }_{0\mu} 
 \bigr] 
 \nn[2mm] 
 & + & 
 \gE^4 
 \bigl[
  c^{ }_{10} A_0^6 \,
 \bigr]
 \Bigr\}
 \;. \hspace*{6mm} \la{chapman}
\ea
The colour trace refers to the adjoint representation: 
$ 
 \mathop{\rm tr}\{ 
                   A B
                \}
 \equiv A^{ }_{ab} B^{ }_{ba}
$, 
$ 
 \mathop{\rm tr}\{ 
                   A B C
                \}
 \equiv A^{ }_{ab} B^{ }_{bc} C^{ }_{ca}
$, 
where 
$(A^{ }_0)^{ }_{ab} \equiv - i f^{abc} A_0^c$, 
$(F^{ }_{\mu 0})^{ }_{ab} \equiv - i f^{abc} F_{\mu 0}^c$, and 
$(D^{ }_\mu F^{ }_{\mu \nu})^{ }_{ab} \equiv - i f^{abc} 
 \mathcal{D}^{cd}_\mu F_{\mu \nu}^d$.
The value of the sum-integral over $P$ evaluates to 
\be
 \Tint{P}' \frac{1}{P^6}
 \; = \;  
 \frac{\Gamma(3-\frac{d}{2})\zeta(6-d)T}{(4\pi)^{\frac{d}{2}}(2\pi T)^{6-d}}
 \; \stackrel{3-2\epsilon}{=} \; 
 \frac{\zeta^{ }_3 \, \mu^{-2\epsilon} }{128 \pi^4 T^2}
 \biggl\{
  1 + 2 \epsilon \biggl[ 
   \ln \biggl( \frac{\bmu e^{\gammaE}}{4\pi T} \biggr) 
 + 1 - \gammaE
 + \frac{\zeta'_3}{\zeta^{ }_3}
 \biggr] 
  + \mathcal{O}(\epsilon^2)
 \biggr\} 
 \;. \la{Ib3}
\ee 

The values of $c^{ }_i$ 
were given for $d=3$ in ref.~\cite{chapman}. 
We need to generalize the expressions to $d$ dimensions, because
some of the operators lead to divergent loop integrals at the second
stage of our analysis (cf.\ \se\ref{se:scalepiT}).
Beyond leading order, the coefficients are also functions of~$g^2$, 
but these contributions are of higher order than the effects
that we are interested in. As mentioned in \se\ref{ss:SEQCD}, 
we are also suppressing effects proportional to $\Nf^{ }$.

As a first step, it may be realized that 
the operator basis in 
\eq\nr{chapman} is redundant: it can be verified that
\be
 \int_X
 \mathop{\rm tr} \Bigl\{ 
 i\gE^{ } \Bigl[ 
   F^{ }_{0 \mu} F^{ }_{\mu\nu} F^{ }_{\nu 0  } + 
   A^{ }_0 (D^{ }_{\mu} F^{ }_{\mu\nu}) F^{ }_{ 0 \nu}  
 \Bigr] 
 + 
 \frac{\gE^2}{2}  
 \Bigl[ 
  - \, A_0^2 F_{\mu\nu}^2 + 
  A^{ }_0 F^{ }_{\mu\nu} A^{ }_0 F^{ }_{\mu\nu} 
 \Bigr]
 \Bigr\} \; = \; 0
 \;.
\ee
Therefore a simultaneous change of the coefficients
($c_i^\rmi{new} \equiv c^{ }_i + \delta c^{ }_i$, $i=4,...,7$) 
has no physical effect, 
provided that
\be
 \delta c^{ }_4 = \delta c^{ }_5 = - 2 \delta c^{ }_6 = 2 \delta c^{ }_7
 \;. \la{Theta_c}
\ee
In particular, we could tune
$c^{ }_7$ to zero as was done in ref.~\cite{chapman},\footnote{%
 Tuning $c^{ }_5$ to zero would yield \eq\nr{chapman} more elegant
 and simplify a number of subsequent computations.  
 } 
by choosing
$\delta c^{ }_7 = - c^{ }_7$. Then \eq\nr{Theta_c} implies that the 
other coefficients should appear in the combinations
\be
 c_4^\rmi{(new)} = c^{ }_4 - 2 c^{ }_7 \;, \quad
 c_5^\rmi{(new)} = c^{ }_5 - 2 c^{ }_7 \;, \quad
 c_6^\rmi{(new)} = c^{ }_6 + c^{ }_7 \;. 
 \la{invariant}
\ee
In the following
we keep both 
$c^{ }_5\neq 0$ and $c^{ }_7\neq 0$
for generality; 
this offers for a good crosscheck in that only the combinations
of \eq\nr{invariant} appear in any physical expressions.

In order to determine the values of the coefficients $c^{ }_i$, we have 
computed 1-loop contributions to 
the 2-point, 3-point, 5-point and 6-point 
functions of the Matsubara zero modes
in the background field Feynman gauge~\cite{lfa}.\footnote{%
 In a general gauge, several of the coefficients
 depend on the gauge fixing parameter,
 but we have checked that the logarithmic divergences
 that we are ultimately interested in do not. 
 } 
Salient details from this computation are presented in
\se\ref{ss:determine}. Matching the 2 and 3-point vertices yields
\be
 c^{ }_1 \; = \; \frac{41-d}{120}
 \;, \quad
 c^{ }_2 \; = \; \frac{(d-1)(d-5)}{120}
 \;, \quad
 c^{ }_3 \; = \; \frac{1-d}{180}
 \;, \quad
 c^{ }_5 - c^{ }_4 \; = \; \frac{(d-1)(d-5)}{60}
 \;. \la{c1}
\ee
Adding the 5-point vertex 
permits for us to fix the combinations in 
\eq\nr{invariant} as  
\be
 c^{ }_4 - 2 c^{ }_7 \; {=} \; \frac{(41-d)(5-d)}{60}
 \;, \quad
 c^{ }_5 - 2 c^{ }_7 \; {=} \; \frac{(21-d)(5-d)}{30}
 \;, \quad
 c^{ }_6 + c^{ }_7 \; {=} \; \frac{(d-25)(5-d)}{24}
 \;. \la{c4}
\ee
In addition the 5-point vertex shows the presence of
so-called evanescent operators whose coefficients
vanish for $d=3$, 
\be 
 c^{ }_8 \; = \; \frac{(5-d)(3-d)(d-1)}{20} \;, \quad
 c^{ }_9 \; = \; \frac{(5-d)(3-d)(d-1)}{30}
 \;. \la{c8}
\ee
The coefficient $c^{ }_{10}$ is also evanescent
and can be determined from 
the 6-point vertex; we find 
$
 c^{ }_{10} = {(5-d)(3-d)(d-1)^2} / {180}
$
but this does not contribute to any of our results. 
For $d=3$ \eqs\nr{c1}--\nr{c8}
agree with ref.~\cite{chapman}. 
(Expressions for a general $d$ were derived in 
ref.~\cite{salcedo}, but unfortunately a rather 
different notation was employed.) 

%
\subsection{Details on the determination of dimension-six coefficients}
\la{ss:determine}

In this section we provide some details on 
the determination of the coefficients listed in \eqs\nr{c1}--\nr{c8}.
The derivation of \eq\nr{chapman} 
is most conveniently formulated with the background
field method~\cite{lfa}, and as a reminder  
the gauge potentials are denoted by $B^a_\mu$. The object
computed is the background field effective action, 
$\Gamma^{ }_\rmii{EQCD}[B]$, whereby the vertices 
are automatically symmetrized in the appropriate way. 
After a field redefinition, {\it viz.}
$A^a_i = B^a_i (1+ \rmO(\gB^2))$ and   
$A^a_0 = B^a_0 (1+ \rmO(\gB^2))$,  
the result is identified with $S^{ }_\rmii{EQCD}[A]$.

We choose to work directly in momentum space, 
with the background fields
denoted by $B^a_\mu({q})$. The momenta ${q}$ have 
spatial components only:
\be
 q^{ }_\mu \equiv \delta^{ }_{\mu i}\, q^{ }_i \;. 
\ee
Specific
tensors are defined for showing the dependence of the vertices on 
spacetime and colour indices; these are summarized in appendix~A.
The structure naturally emerging from the computation is one in 
which there are Lorentz-invariant structures ($\delta^{ }_{\mu\nu}$ etc.)
and additional terms that only appear for the zero components of the
gauge potentials; the latter are identified through the 
tensors $T^{ }_{\mu\nu} \equiv \delta^{ }_{\mu 0}\delta^{ }_{\nu 0}$ etc.
Results for various 1-loop sum-integrals in this basis are given
in appendix~B.

Computing the 2-point and 3-point vertices in the background field
gauge, we obtain the 1-loop correction 
\ba
 \Gamma^{(2+3)}_\rmii{EQCD}[B] 
 & = &  \frac{\gB^2 \Nc^{ }}{2!}\, B^a_\mu(q)\, B^b_\nu(r) 
 \, \delta^{ab} \, \delta(q+r)  
 \, \gamma^{(2)}_{\mu\nu}(q)
 \nn 
 & + & 
 \frac{  i \gB^3 \Nc^{ } }{3!}\, B^a_\mu(q)\, B^b_\nu(r)\, B^c_\rho(s)
 \, f^{abc} \, \delta(q+r+s)  
 \, \gamma^{(3)}_{\mu\nu\rho}(q,r,s)
 \;, \hspace*{6mm} \la{L_EQCD_2_3}
\ea
where summations and integrations are implied, 
and $T \int_q \delta(q) \equiv 1$. 
Expanding in $1/P^2 \sim 1/(\pi T)^2$,
the 2-point vertex reads
\ba
 \gamma^{(2)}_{\mu\nu}(q) & = & 
 \Tint{P}' \biggl\{ 
   \frac{(d-25)
   \bigl( q^2 \delta^{ }_{\mu\nu} - q^{ }_\mu  q^{ }_\mu \bigr)
   }{6 P^4} 
  + 
   T^{ }_{\mu \nu} \,
   \biggl[ 
    \frac{(d-1)^2}{P^2} - \frac{(d-1)(d-3) q^2}{6 P^4}
   \biggr]
  \nn & + & 
   \frac{4 c^{ }_1 \, q^2
   \bigl( q^2 \delta^{ }_{\mu\nu} - q^{ }_\mu  q^{ }_\mu \bigr)
  +  
  4 c^{ }_2 \, q^4\, T^{ }_{\mu \nu} }{P^6}
  \; + \;   \rmO\Bigl( \frac{1}{P^8} \Bigr)
 \biggr\}
 \;, \quad \la{2pt}
\ea
where $c^{ }_1$ and $c^{ }_2$ have the values in \eq\nr{c1}. 
The term proportional to $\Tintip{P} \frac{1}{P^2}$ yields 
the parameter $\mE^2$ in \eq\nr{mE}, whereas the terms 
proportional to $\Tintip{P} \frac{1}{P^4}$ yield wave
function corrections. The existence of 
a term $\Tintip{P} \frac{T^{ }_{\mu\nu} q^2}{P^4}$ indicates
that temporal and spatial components of the gauge 
potentials need to be normalized differently. 

For the 3-point vertex a similar computation leads to 
\ba
 & & \hspace*{-1cm} \gamma^{(3)}_{\mu\nu\rho}(q,r,s)  =  
 \Tint{P}' \biggl\{ 
   \frac{(25 - d)
   q^{ }_\rho \delta^{ }_{\mu\nu} 
     + (d-1)(d-3)\,
     q^{ }_\rho T^{ }_{\mu\nu} }{P^4}
  \nn & - & 
  \frac{ 24 c^{ }_1\, q^{ }_{\mu} q^{ }_{\rho} r^{ }_{\nu}
  + 12 c^{ }_3\, q^{ }_\nu
  ( r^{ }_\mu  q^{ }_\rho - q^{ }_{\mu}  r^{ }_\rho )  
  }{P^6}
  \nn & - &  
   \frac{
    6(4c^{ }_1-3c^{ }_3) s^2 q^{ }_\rho \, \delta^{ }_{\mu \nu} 
   - 6 q^2 [ 3 c^{ }_3\,  s^{ }_\rho + 8 c^{ }_1\,  r^{ }_\rho   ] 
   \, \delta^{ }_{\mu \nu} 
   }{P^6}
 \nn & + &  
    \frac{6(c^{ }_4 - c^{ }_5) s^2\, q^{ }_\rho    T^{ }_{\mu \nu} \, 
   - 6 q^2 [ 4 c^{ }_2  (q^{ }_\rho - r^{ }_\rho) 
   +  (c^{ }_5 - c^{ }_4) s^{ }_\rho ]    T^{ }_{\mu \nu} \,
     }{P^6}
  \; + \; \rmO\Bigl( \frac{1}{P^8} \Bigr)
 \biggr\}
 \;, \la{3pt}
\ea
where $c^{ }_3$ and $c^{ }_4 - c^{ }_5$ have the values shown 
in \eq\nr{c1}.\footnote{%
 This representation is not unique, cf.\ the 
 comments below \eq\nr{3pt_ops}.
 }
The terms proportional to $\Tintip{P} \frac{1}{P^4}$ can be 
partly accounted for by wave function corrections; the remainder
yields the effective gauge coupling of \eq\nr{gE}. 
The same result for $\gE^2$
is obtained both from a purely spatial vertex 
($\sim q^{ }_\rho \delta^{ }_{\mu i}\delta^{ }_{\nu i}$)
 and from a vertex mixing two $A^a_0$'s and one $A^a_i$
($\sim q^{ }_\rho T^{ }_{\mu\nu}$).

The 4-point vertex can similarly be written as 
\be
 \Gamma^{(4)}_\rmii{EQCD}[B] 
 \; = \;  \frac{\gB^4 }{4!}\, B^a_\mu(q)\, B^b_\nu(r) \,
 B^c_\alpha(s)\, B^d_\beta(t)   
 \, \delta(q+r+s+t) 
 \, \gamma^{(4)abcd}_{\mu\nu\alpha\beta}(q,r,s,t)
 \;, 
\ee
where 
\ba
 & & \hspace*{-1cm} \gamma^{(4)abcd}_{\mu\nu\alpha\beta}(q,r,s,t)  =  
 \Tint{P}' \biggl\{ 
   X^{\{ab\}\{cd\}} \, \frac{2 (d-1)^2(3-d)  T^{ }_{\mu\nu\alpha\beta}}{P^4}
 \nn 
 & + & 
    X^{[ab][cd]} \, \frac{4(25-d)  
      \delta^{ }_{\mu\alpha} \delta^{ }_{\nu\beta} 
     + 8 (d-1)(d-3) T^{ }_{\mu\alpha} \delta^{ }_{\nu\beta}}{P^4}
 +  \rmO\biggl( \frac{1}{P^6} \biggr) \biggr\} 
 \;. \hspace*{5mm} \la{sc_4_pre}
\ea
The notations 
$ X^{\{ab\}\{cd\}} $ 
and 
$ X^{[ab][cd]} $ 
are defined in appendix~A. 
The term proportional to $\Tintip{P} \frac{T^{ }_{\mu\nu\alpha\beta}}{P^4}$
yields $\lE^{ }$ in \eq\nr{lE}, whereas 
the other terms proportional to $\Tintip{P} \frac{1}{P^4}$ 
correspond to wave function corrections and $\gE^2$.
The dimension-six part of the 4-point vertex 
is rather complicated
(it is shown in appendix~C) and we have 
not used it for determining $c^{ }_i$'s.

%
\begin{figure}[t]
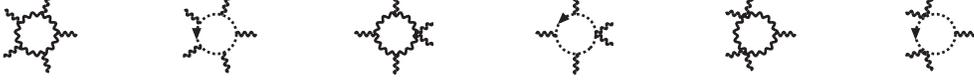


\hspace*{0.2cm}%
\begin{minipage}[c]{14cm}
\begin{eqnarray*}
 && 
 \DiagA \qquad 
 \DiagB \qquad
 \DiagC \qquad
 \DiagD \qquad
 \DiagE \qquad
 \DiagF 
\end{eqnarray*}
\end{minipage}

\caption[a]{\small 
1-loop contributions to the 5-point function 
in the background field gauge.
Wiggly lines denote gluons and dotted lines ghosts. 
The diagrams have been drawn with Axodraw~\cite{axodraw}.
}
\la{fig:5pt}
\end{figure}
%

Proceeding finally to the 5-point vertex, we find 
no contribution $\sim \Tintip{P} \frac{1}{P^4}$. 
The contribution of the dimension-six operators
from \eq\nr{chapman} can be written as
\ba
 \Gamma^{(5)}_\rmii{EQCD}[B] 
 & = & 
 B^a_\mu(q)\, B^b_\nu(r) \, B^c_\rho(s) \, 
 B^d_\alpha(t)\, B^e_\beta(u)   
 \, \delta(q+r+s+t+u) 
 \, \biggl( \Tint{P}' \frac{ 8 i \gE^5 s^{ }_\mu }{P^6} \biggr)
 \nn 
 & \times &  
 \Bigl\{ 
 X^{\{ab\}[cde]}\,
 \Bigl[
    - c^{ }_1 \, \delta^{ }_{\rho\alpha} \delta^{ }_{\nu\beta}
    +4 c^{ }_1\,  \delta^{ }_{\rho\beta} \delta^{ }_{\nu\alpha}
    - c^{ }_1 \, \delta^{ }_{\rho\nu} \delta^{ }_{\alpha\beta}
 \nn 
 & & \hspace*{18mm}
 -  c^{ }_2 \,  T^{ }_{\rho\alpha} \delta^{ }_{\nu\beta}
 + 4 c^{ }_2\,  T^{ }_{\rho\beta} \delta^{ }_{\nu\alpha}
 -  c^{ }_2\,  T^{ }_{\rho\nu} \delta^{ }_{\alpha\beta} 
 \nn 
 & & \hspace*{18mm}
  - c^{ }_2 \, \delta^{ }_{\rho\alpha} T^{ }_{\nu\beta}
 + (c^{ }_5 - 2 c^{ }_7) \, \delta^{ }_{\rho\beta} T^{ }_{\nu\alpha} 
 -  c^{ }_2\, \delta^{ }_{\rho\nu} T^{ }_{\alpha\beta}
 -  c^{ }_9  T^{ }_{\rho\nu\alpha\beta}
 \Bigr]
 \nn 
 & + & 
 X^{[ab]\{cde\}}\,
 \Bigl[
   (5 c^{ }_1 - 3 c^{ }_3)\, \delta^{ }_{\rho\alpha} \delta^{ }_{\nu\beta}
 +\, (3 c^{ }_3 - 4c^{ }_1)\, \delta^{ }_{\rho\beta} \delta^{ }_{\nu\alpha}
 +\, 3 c^{ }_1\, \delta^{ }_{\rho\nu} \delta^{ }_{\alpha\beta}
 \nn 
 & & \hspace*{16mm}
 +  (c^{ }_2 - c^{ }_4 + c^{ }_5 ) \,  T^{ }_{\rho\alpha} \delta^{ }_{\nu\beta}
 +  (c^{ }_4 - c^{ }_5) \,  T^{ }_{\rho\beta} \delta^{ }_{\nu\alpha}
 + 3 c^{ }_2 \, T^{ }_{\rho\nu} \delta^{ }_{\alpha\beta}  
 \nn 
 & & \hspace*{16mm}
 +  (c^{ }_2  - c^{ }_4 + c^{ }_5)
   \, \delta^{ }_{\rho\alpha} T^{ }_{\nu\beta}
 +  (c^{ }_4 -4 c^{ }_2 - 2 c^{ }_7)
   \, \delta^{ }_{\rho\beta} T^{ }_{\nu\alpha}
 \nn 
 & & \hspace*{16mm}
 +\,  (c^{ }_5 -c^{ }_2 + 2 c^{ }_6)
  \, \delta^{ }_{\rho\nu} T^{ }_{\alpha\beta}
 +  (c^{ }_8 - c^{ }_9) \, T^{ }_{\rho\nu\alpha\beta}  
 \Bigr] \, 
 \Bigr\} 
 + 
 \rmO\biggl(  
 \Tint{P}' \frac{1}{P^8}
 \biggr)
 \;. \hspace*{8mm}
 \la{5pt}
\ea
We have computed the corresponding Feynman diagrams, 
shown in \fig\ref{fig:5pt}. Making use of momentum conservation
and appropriate symmetrizations, and identifying
$\gE^2 = \gB^2 (1 + \rmO(\gB^2))$, 
we obtain precisely the same structure from Feynman diagrams. 
There are 20 independent terms 
that permit for a crosscheck of \eq\nr{c1} and, most importantly, 
for a unique determination of the combinations appearing 
in \eqs\nr{c4} and \nr{c8}. 

%
\section{Overlapping soft/hard and ultrasoft/hard contributions}
\la{se:scalepiT}

In EQCD, the gauge field components $A_0^a$ have turned
into massive adjoint scalar fields when the non-zero Matsubara
modes were integrated out (cf.\ \eq\nr{S_EQCD}). Our goal now
is to integrate out the massive $A_0^a$, 
and thereby construct the MQCD action.
Its super-renormalizable part has the form of the spatial part of 
\eq\nr{S_EQCD}. We denote it by 
\be
 S^{ }_\rmii{MQCD}[A] \; \equiv \; 
 \int_X 
  \fr14 F^a_{ij}F^a_{ij}
 \;, \la{S_MQCD}
\ee
even though $ F^a_{ij} $ now contains a different gauge coupling 
than \eq\nr{S_EQCD}: 
$
 F^a_{ij} = \partial^{ }_i A^a_j - \partial^{ }_j A^a_i
 + \gM^{ } f^{abc} A^b_i A^c_j 
$.
The main goal of this section is to determine the contributions
to $\gM^2$ that originate from the dimension-six operators 
in \eq\nr{chapman}. These are termed soft/hard 
(\ses\ref{ss:1loop} and \ref{ss:2loop}) and ultrasoft/hard
(\se\ref{ss:MQCD}) contributions. 

We note 
that in analogy with \eq\nr{chapman}, $S^{ }_\rmii{MQCD}$ also 
has a dimension-six part, $\delta S^{ }_\rmii{MQCD}$. It is given 
in \eq\nr{chapman_MQCD} and discussed in more detail 
in \se\ref{ss:MQCD}.

In order to determine $\gM^2$, we once again make use of the 
background field effective action, $\Gamma^{ }_\rmii{MQCD}[B]$.
In particular, we consider its quadratic part, 
\be
 \Gamma^{(2)}_\rmii{MQCD} [B]  = 
 \fr12 B^a_i(q)\, B^a_j(-q)
 (q^2 \delta^{ }_{ij} - q^{ }_i q^{ }_j )\,
 \bigl( Z^{ }_{\bg} + \delta Z^{ }_{\bg} \bigr)
 \;, \la{G_MQCD_2}
\ee
where
$
 \delta Z^{ }_{\bg}
$
collects any possible divergences. 

In the background field gauge, $\Gamma$ is gauge invariant in terms
of $B$~\cite{lfa}.  
Consequently the 3-point and 4-point vertices are fully determined 
by \eq\nr{G_MQCD_2}. After a subsequent field redefinition, this 
implies that $Z^{ }_{\bg}$ determines the gauge coupling of MQCD: 
\be 
 \gM^2 = \gER^2\, \mu^{2\epsilon} \, Z_{\bg}^{-1} 
 - \gER^2\, \mu^{2\epsilon} \, \delta Z_{\bg}^{ } 
 + \delta \gE^2 + \rmO(g^{10})
 \;. 
\ee
Here $ \delta \gE^2 $ is from \eq\nr{gE_ct}. 
The following discussion is carried out in terms 
of $Z^{ }_{\bg}$ and $\delta Z^{ }_{\bg}$. 

When the field $A_0^a$ is integrated out and
one vertex from \eq\nr{chapman} is included, 
we expect to find terms of the types 
\be
 Z^{ }_{\bg} + \delta Z^{ }_{\bg} = 
 1 + \biggl(  \Tint{P}'\frac{\gE^2 \Nc^{ }}{P^6} \biggr) 
 \, \biggl[ \frac{\mER^{ } \gER^2 \Nc^{ } T}{4\pi}
 \, \#^{(5)}_{ }
 + 
 \,  \frac{(\gER^2\Nc^{ } T)^2}{(4\pi)^2}
 \,  \#^{(6)}_{ } 
 + \ldots \biggr] \;, \la{dZ2}
\ee
where $\#^{(6)}_{ } $ may contain logarithms. 
The corresponding effects are  of 
$\rmO(g^5)$ and $\rmO(g^6)$ in terms of the original QCD coupling. 
The latter effect is comparable to \eq\nr{ioan_eps}. 

Before proceeding let us explain why we consider
``2-loop soft $\times$ 1-loop hard'' contributions,
i.e.\ 2-loop graphs with one insertion of dimension-six operators, 
but not ``1-loop soft $\times$ 2-loop hard'' ones. 
In terms of $Z^{ }_{\bg}$ defined in \eq\nr{G_MQCD_2}, 
``1-loop hard'' gives a factor $\sim g^2/T^2$, 
``1-loop soft'' gives a factor $\sim g^2 T \mER^{ }\sim g^3 T^2$, and
``2-loop soft'' gives a factor $\sim (g^2 T)^2\sim g^4 T^2$. 
The overall effects of these orders are 
$\sim g^5, g^6$, cf.\ \eq\nr{dZ2}. 
In contrast ``2-loop hard'' would give dimension-six operators
proportional to $\sim g^4/T^2$. The overall effect from 
``1-loop soft $\times$ 2-loop hard'' would therefore
be $\sim g^7$, i.e.\ of higher order  
than our computation. The same applies to 
dimension-eight operators, whose coefficients are 
$\sim g^2/T^4$ and who get a further suppression factor 
$\lsim g^2 T \mER^3 \sim g^5 T^4$ from soft effects.

%
\subsection{1-loop results with dimension-six operators}
\la{ss:1loop}

%
\begin{figure}[t]
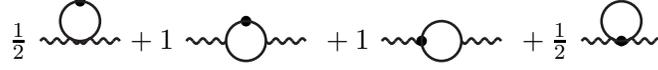


\hspace*{2.5cm}%
\begin{minipage}[c]{10cm}
\begin{eqnarray*}
 && 
 \tfr{1}{2} \ScatA \hspace*{-4mm} +  
 {\textstyle{1}}\;\;\; \ScatB + 
 {\textstyle{1}}\;\;\; \ScatD + 
 \tfr{1}{2} \ScatC 
\end{eqnarray*}
\end{minipage}

\caption[a]{\small 
1-loop contributions to the 2-point function, 
containing some of the ``Chapman vertices'' from \eq\nr{chapman}, 
denoted by a filled blob. The 
adjoint scalar fields are denoted by solid lines.  
}
\la{fig:1loop}
\end{figure}
%

The 1-loop contribution to $Z^{ }_{\bg}$ from dimension-six
operators originates from 
the graphs shown in \fig\ref{fig:1loop}.  
The vertices related to dimension-six operators have been
indicated with a filled blob; we refer to them 
as ``Chapman vertices''. 
In appendix~C the vertices are written 
in a form convenient for computing these graphs. 
The 2-point vertex is parametrized through
$\varzeta^{ }_1,\varzeta^{ }_2$, cf.\ \eq\nr{2pt_ops}; 
the 3-point vertex through $\xi^{ }_{1},...,\xi^{ }_{10}$, 
cf.\ \eq\nr{3pt_ops}; 
and the 4-point vertex through $\psi^{ }_{1},...,\psi^{ }_{44}$ 
and $\omega^{ }_{1},...,\omega^{ }_{35}$, cf.\ \eq\nr{4pt_ops}. 

Computing the graphs in \fig\ref{fig:1loop} in dimensional regularization
and expanding in $q^2/\mE^2$,
all of them can be related to a single 1-loop tadpole integral, denoted by
\be
 I(\mE^{ }) \; \equiv \; 
 \int_{p} \frac{T}{p^2 + \mE^2} 
 \; = \; 
 \frac{\mE^{d-2}\Gamma(1-\frac{d}{2})T}{(4\pi)^{\frac{d}{2}}}
 \; \stackrel{3-2\epsilon}{=} \; 
 - \frac{ \mE^{ } T \mu^{-2\epsilon}}{4\pi}\, 
 \biggl[ 1 + 2\epsilon\,\Bigl( 1 + \ln\frac{\bmu}{2 \mE^{ }}\Bigr)  
 + \rmO(\epsilon^2) \biggr]
 \;. \la{I}
\ee
We get 
\ba
 \delta^{ } \Gamma^{(2)}_\rmii{MQCD} [B] & = & 
 B^a_i(q)\, B^b_j(r)  \, \delta^{ab} \, \delta(q+r) \, 
 \biggl(  \Tint{P}'\frac{\gE^4 \Nc^2}{P^6} \biggr)
 \, I(\mE^{ }) \, 
 \nn & \times & 
 \biggl\{
 \mE^2\, \delta^{ }_{ij} 
 \biggl[ 
    \frac{d+2}{d} \, \bigl( -2\varzeta^{ }_2 - \xi^{ }_8 + \xi^{ }_9 \bigr) 
  - \fr34 \Bigl( \psi^{ }_4 + \frac{\psi^{ }_{26}}{d} \Bigr) 
 \nn & & \hspace*{1cm}
  -\, \psi^{ }_{13}
  + \psi^{ }_{15}
  + \frac{1}{d} \bigl( \psi^{ }_{35}  - \psi^{ }_{34}  \bigr)
  + \fr14 \Bigl(\omega^{ }_4 + \frac{\omega^{ }_{26}}{d}  \Bigr)
 \biggr]
 \nn & + & 
 \bigl(q^2 \delta^{ }_{ij} - q^{ }_i q^{ }_j \bigr) 
 \biggl[ 
    \frac{(4+d)(2-d)}{24}\, \varzeta^{ }_2 
  + \frac{d-2}{12} \, \bigl( \xi^{ }_9 - \xi^{ }_8 \bigr) + \xi^{ }_{10}
 \nn & & \hspace*{1cm}
  + \, \frac{3 \psi^{ }_5}{4}  
  + \psi^{ }_{16}  - \psi^{ }_{18} 
  - \frac{\omega^{ }_5}{4} 
 \biggr]
 \nn & + & 
 q^{ }_i q^{ }_j \, 
 \biggl[ 
    \varzeta^{ }_2 
  +  \xi^{ }_8 + \xi^{ }_{10}
  + \fr34 \bigl( \psi^{ }_5 + \psi^{ }_{29}\bigr)
  + \psi^{ }_{16}  - \psi^{ }_{18}
 \nn & & \hspace*{1cm} 
  +\, \psi^{ }_{42} - \psi^{ }_{43} - \psi^{ }_{44} 
  - \fr14 (\omega^{ }_5 + \omega^{ }_{29})
 \biggr] 
 \; + \; \rmO\Bigl( \frac{q^4}{\mE^2} \Bigr)
 \biggr\}
 \;. \la{MQCD_1l}
\ea
Inserting the values of the coefficients in terms of the $c^{ }_i$'s from 
appendix~C, the terms proportional 
to $\mE^2 \delta^{ }_{ij}$ and $q^{ }_i q^{ }_j$ drop out
as required by gauge invariance, and we are left with
\ba
 \delta^{ } \Gamma^{(2)}_\rmii{MQCD} [B] & = & 
 B^a_i(q)\, B^b_j(r)  \, \delta^{ab} \, \delta(q+r) \, 
 (q^2 \delta^{ }_{ij} - q^{ }_i q^{ }_j ) 
 \, \biggl(  \Tint{P}'\frac{\gE^4 \Nc^2}{P^6} \biggr)
 \, I(\mE^{ }) \, 
 \nn & \times & 
 \biggl\{ 
  \frac{(4-d)(d-2)}{12}\, (c^{ }_1 + c^{ }_2)
  + 3 c^{ }_3 + (c^{ }_4 - 2 c^{ }_7 ) + 4 (c^{ }_6 + c^{ }_7) 
 \biggr\}
 \;. \la{1loopgM}
\ea
Inserting the coefficients $c^{ }_1,...,c^{ }_7$ 
from \eqs\nr{c1} and \nr{c4} and setting $d\to 3$, 
the curly brackets evaluate to 
\ba
 \lim_{d\to 3}
 \bigl\{ ... \bigr\} & = & 
 - \lim_{d\to 3}
 \frac{d^4-13\, d^3+312\, d^2 - 6404\, d + 25424}{1440}
 \; = \; -\frac{875}{144}
 \;. \la{cZ5}
\ea
The corresponding contribution to $Z^{ }_{\bg}$ is shown 
on the first row of \eq\nr{finalZ2}. 

%
\subsection{2-loop results with dimension-six operators}
\la{ss:2loop}

%
\begin{figure}[t]
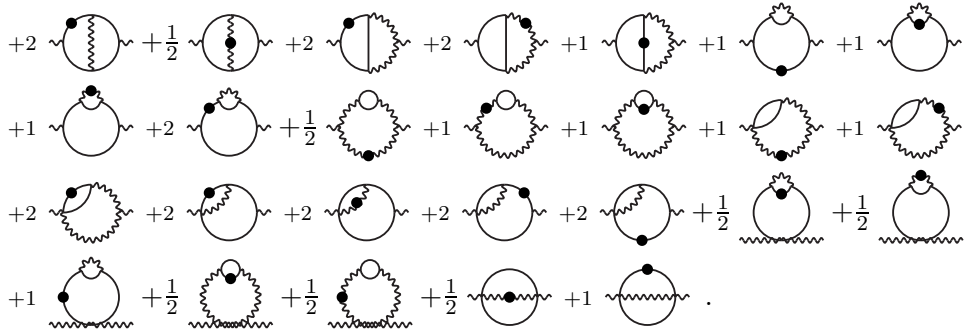


\hspace*{0.1cm}%
\begin{minipage}[c]{15.0cm}
\begin{eqnarray*}
&&{} \hspace*{-0.3cm}
\sm{+2} \ToptSMx(\Legl,\Asc,\Asc,\Asc,\Asc,\Lgl) 
\sy+12  \ToptSMy(\Legl,\Asc,\Asc,\Asc,\Asc,\Lgl) 
\sm{+2} \ToptSMx(\Legl,\Agl,\Asc,\Asc,\Agl,\Lsc)
\sm{+2} \ToptSMz(\Legl,\Agl,\Asc,\Asc,\Agl,\Lsc)
\sm{+1} \ToptSMy(\Legl,\Agl,\Asc,\Asc,\Agl,\Lsc)
\sm{+1} \ToprSBBy(\Legl,\Asc,\Asc,\Agl,\Asc)
\sm{+1} \ToprSBBz(\Legl,\Asc,\Asc,\Agl,\Asc)
 \nn[1.5ex]&&{} \hspace*{-0.3cm}
\sm{+1} \ToprSBBw(\Legl,\Asc,\Asc,\Agl,\Asc)
\sm{+2} \ToprSBBx(\Legl,\Asc,\Asc,\Agl,\Asc)
\sy+12  \ToprSBBy(\Legl,\Agl,\Agl,\Asc,\Asc)
\sm{+1} \ToprSBBx(\Legl,\Agl,\Agl,\Asc,\Asc)
\sm{+1} \ToprSBBz(\Legl,\Agl,\Agl,\Asc,\Asc)
\sm{+1} \ToptSAly(\Legl,\Agl,\Asc,\Agl,\Asc)
\sm{+1} \ToptSAlz(\Legl,\Agl,\Asc,\Agl,\Asc)
 \nn[1.5ex]&&{} \hspace*{-0.3cm}
\sm{+2} \ToptSAlx(\Legl,\Agl,\Asc,\Agl,\Asc)
\sm{+2} \ToptSAlx(\Legl,\Asc,\Asc,\Asc,\Agl)
\sm{+2} \ToptSAlw(\Legl,\Asc,\Asc,\Asc,\Agl)
\sm{+2} \ToptSAlz(\Legl,\Asc,\Asc,\Asc,\Agl)
\sm{+2} \ToptSAly(\Legl,\Asc,\Asc,\Asc,\Agl)
\sy+12  \ToprSTBy(\Legl,\Asc,\Agl,\Asc)
\sy+12  \ToprSTBz(\Legl,\Asc,\Agl,\Asc)
 \nn[1.5ex]&&{} \hspace*{-0.3cm}
\sm{+1} \ToprSTBx(\Legl,\Asc,\Agl,\Asc)
\sy+12  \ToprSTBy(\Legl,\Agl,\Asc,\Asc) 
\sy+12  \ToprSTBx(\Legl,\Agl,\Asc,\Asc) 
\sy+12  \ToptSSx(\Lgl,\Asc,\Asc,\Lgl) 
\sm{+1} \ToptSSy(\Lgl,\Asc,\Asc,\Lgl) 
 \;.
\end{eqnarray*}
\end{minipage}

\vspace*{5mm}

\caption[a]{\small 
2-loop contributions to the 2-point function, 
originating from 2-point Chapman vertices, denoted by filled blobs. 
Adjoint scalars are denoted by solid lines. Graphs involving closed massless
loops, which do not contribute to the matching, have been omitted.  
}
\la{fig:2loop2}
\end{figure}
%

%
\begin{figure}[t]

\hspace*{0.1cm}%
\begin{minipage}[c]{15.0cm}
\begin{eqnarray*}
&&{} \hspace*{-0.3cm}
\sm{+1} \ToptSMa(\Legl,\Asc,\Asc,\Asc,\Asc,\Lgl) 
\sm{+1} \ToptSMb(\Legl,\Asc,\Asc,\Asc,\Asc,\Lgl) 
\sm{+1} \ToptSMa(\Legl,\Agl,\Asc,\Asc,\Agl,\Lsc)
\sm{+2} \ToptSMb(\Legl,\Agl,\Asc,\Asc,\Agl,\Lsc)
\sm{+2} \ToprSBBa(\Legl,\Asc,\Asc,\Agl,\Asc)
\sm{+2} \ToprSBBb(\Legl,\Asc,\Asc,\Agl,\Asc)
\sm{+1} \ToprSBBa(\Legl,\Agl,\Agl,\Asc,\Asc)
 \nn[1.5ex]&&{} \hspace*{-0.3cm}
\sm{+1} \ToprSBBb(\Legl,\Agl,\Agl,\Asc,\Asc)
\sm{+1} \ToptSAla(\Legl,\Agl,\Asc,\Agl,\Asc)
\sm{+1} \ToptSAlb(\Legl,\Agl,\Asc,\Agl,\Asc)
\sm{+2} \ToptSAla(\Legl,\Asc,\Asc,\Asc,\Agl)
\sm{+2} \ToptSAlb(\Legl,\Asc,\Asc,\Asc,\Agl)
\sm{+1} \ToprSTBa(\Legl,\Asc,\Agl,\Asc)
\sy+12  \ToprSTBa(\Legl,\Agl,\Asc,\Asc) 
\end{eqnarray*}
\end{minipage}

\vspace*{5mm}

\caption[a]{\small 
2-loop contributions to the 2-point function, 
originating from 3-point Chapman vertices (the notation is as 
in \fig\ref{fig:2loop2}). 
}
\la{fig:2loop3}
\end{figure}
%

%
\begin{figure}[t]

\hspace*{0.1cm}%
\begin{minipage}[c]{15.0cm}
\begin{eqnarray*}
&&{} \hspace*{-0.3cm}
\sm{+1} \ToptSAlc(\Legl,\Agl,\Asc,\Agl,\Asc)
\sm{+2} \ToptSAlc(\Legl,\Asc,\Asc,\Asc,\Agl)
\sy+12  \ToprSTBc(\Legl,\Asc,\Agl,\Asc)
\sy+14  \ToprSTBc(\Legl,\Agl,\Asc,\Asc) 
\sm{+1} \ToptSSc(\Lgl,\Asc,\Asc,\Lgl) 
\sy+12 \ToprSBTc(\Legl,\Asc,\Asc,\Asc)
\sy+14 \ToprSTTc(\Legl,\Asc,\Asc) 
\sy+14 \ToptSEc(\Lgl,\Asc,\Asc,\Asc,\Asc)
\end{eqnarray*}
\end{minipage}

\vspace*{5mm}

\caption[a]{\small 
2-loop contributions to the 2-point function, 
originating from 4-point Chapman vertices (the notation is as 
in \fig\ref{fig:2loop2}). 
}
\la{fig:2loop4}
\end{figure}
%

%
\begin{figure}[t]

\vspace*{5mm}

\hspace*{2.1cm}%
\begin{minipage}[c]{10cm}
\begin{eqnarray*}
 && 
 + \tfr{1}{2} \FiveA \hspace*{-0.4cm} + 
 \tfr{1}{2}\;\; \FiveB \hspace*{0.0cm} + 
 \tfr{1}{8} \SixA 
\end{eqnarray*}
\end{minipage}

\vspace*{1mm}

\caption[a]{\small 
2-loop contributions to the 2-point function, 
originating from 5-point or 6-point Chapman vertices (the notation is as 
in \fig\ref{fig:2loop2}). 
}
\la{fig:2loop56}
\end{figure}
%

At 2-loop level, 
the contributions of the 2-point, 3-point and 4-point Chapman vertices
to $Z^{ }_{\bg}$ can be extracted from Feynman diagrams
shown in \figs\ref{fig:2loop2}--\ref{fig:2loop4}.
In addition the 5-point and 6-point Chapman 
vertex also contribute. 
The general expressions for these, parametrized through the coefficients
$\kappa_1,...,\kappa_{10}$, $\lambda_1,...,\lambda_{10}$ and 
$\chi_1,...,\chi^{ }_{16}$, are given in \eqs\nr{5pt_ops}
and \nr{6pt_ops}, respectively, and  
the corresponding diagrams are shown in \fig\ref{fig:2loop56}.

In order to display the result, 
we introduce a 2-loop ``sunset'' integral, 
\ba
 H(\mE^{ }) & \equiv &  \int_{{p},{q}} 
 \frac{T^2}{(p^2 + \mE^2)(q^2 + \mE^2)({p+q})^2} \nn
 & = & 
 \frac{\mE^{2d-6} \Gamma(1-\frac{d}{2}) \Gamma(2-\frac{d}{2})  T^2 }
 {(d-3) (4\pi)^{d}}
 \; \stackrel{3-2\epsilon}{=} 
 \; \frac{ T^2 \mu^{-4\epsilon}_{ }}{(4\pi)^2}
 \biggl[ \frac{1}{4\epsilon} + \ln\biggl(\frac{\bmu}{2\mE^{ }}\biggr)
  + \fr12 + \rmO(\epsilon) \biggr]
 \;. \hspace*{5mm} \la{H}
\ea
Then 
\ba
 \delta^{ } \Gamma^{(2)}_\rmii{MQCD} [B] & = & 
 \fr12 B^a_i (q)\, B^b_j (r)  \, \delta^{ab} \, \delta(q+r) \, 
 \, 
 \biggl(  \Tint{P}'\frac{\gE^6 \Nc^3}{P^6} \biggr) \, H(\mE^{ })\, 
 \nn & \times & 
 \biggl\{
  \frac{ \mE^2\, \delta^{ }_{ij} }{4d} \, C^{ }_1 
  \; + \; 
 \frac{ q^2 \delta^{ }_{ij} - q^{ }_i q^{ }_j  }{4d} \, C^{ }_2 
  \; + \; 
 \frac{ q^{ }_i q^{ }_j }{4d} \, C^{ }_3 
 \; + \; \rmO\Bigl( \frac{q^4}{\mE^2} \Bigr)
 \biggr\}
 \;, \hspace*{6mm} \la{2loopgM5}
\ea
where $C^{ }_1,C^{ }_2,C^{ }_3$ are given in appendix~E in terms
of the coefficients 
$\eta^{ }_1, ... ,
\chi^{ }_{16}$.\footnote{%
 A general gauge parameter, denoted by $\alpha$, 
 has been employed: 
 $\langle A^a_k(p) A^b_l(q) \rangle \equiv 
 \frac{\delta^{ab}\delta(p+q)}{p^2} \bigl( \delta^{ }_{kl}
  - \frac{\alpha p^{ }_k p^{ }_l}{p^2}\bigr)$.
 }
 
Inserting the values of the coefficients from appendix~C,  
we find that $C^{ }_1$ and $C^{ }_3$ and terms
proportional to $\alpha$ in $C^{ }_2$ cancel. 
The remaining contribution reads
\ba
 \delta^{ } \Gamma^{(2)}_\rmii{MQCD} [B] & = &
 - B^a_i (q)\, B^b_j (r)  \, \delta^{ab} \, \delta(q+r) \, 
 \bigl(  q^2 \delta^{ }_{ij} - q^{ }_i q^{ }_j  \bigr)
 \, 
 \biggl(  \Tint{P}'\frac{\gE^6 \Nc^3}{P^6} \biggr) \, H(\mE^{ })\, 
 \nn & \times & 
 \biggl\{
  \frac{(d-3)(d-4)^2(d^3-10d^2+23d-44)(c^{ }_1 + c^{ }_2)}{6d(d-5)(d-7)} 
  \nn 
  & + &
  \frac{(d^4 - 18 d^3 + 95 d^2 - 210 d + 192)c^{ }_3}{2d(d-5)}
  \; + \; 
  \frac{(d^3 - 13 d^2 + 36 d - 36)(c^{ }_4 - 2 c^{ }_7)}{6d}
  \nn 
  & + &
  \frac{2(d^3 - 13 d^2 + 21 d - 6)(c^{ }_6 + c^{ }_7)}{3d}
  \; + \; 
  \frac{(d-3)(d-4)(2 c^{ }_8 + c^{ }_9)}{6}
 \biggr\}
  \la{2loopgM5_2} \\[2mm]
 & = &  
 - B^a_i (q)\, B^b_j (r)  \, \delta^{ab} \, \delta(q+r) \, 
 \bigl(  q^2 \delta^{ }_{ij} - q^{ }_i q^{ }_j  \bigr)
 \, 
 \biggl(  \Tint{P}'\frac{\gE^6 \Nc^3}{P^6} \biggr) \,
  H(\mE^{ }) \, 
 \nn & \times & 
 \Bigl(  
    17 d^{\,8} - 494 d^{\,7} + 6522 d^{\,6} - 53766 d^{\,5}
  + 301049 d^{\,4} -  1075772 d^{\,3}
 \nn & & \; 
  + \, 2085956 d^{\,2} - 1575176 d + 102864
 \Bigr)\, \frac{1}{720d(d-5)(d-7)}
 \;,   \la{2loopgM5_3}
\ea
where in the last step we made use of \eqs\nr{c1}--\nr{c8}. 
We note that the evanescent operators parametrized by $c^{ }_8$ and $c^{ }_9$
do not play a role for $d \approx 3$, because the coefficients
with which they contribute in \eq\nr{2loopgM5_2} themselves vanish
for $d\to 3$.

Setting $d=3-2\epsilon$, inserting \eqs\nr{Ib3}, \nr{1loopgM} and \nr{H}, 
and going over to renormalized parameters, we obtain
\ba
 Z^{ }_{\bg} & = & 
 1 \,+\, 
 \biggl( \frac{\gER^2 \Nc^{ }}{16\pi^2 } \biggr)^2
 \frac{ \mER^{ }}{2\pi T}
 \biggl( \frac{875 \zeta^{ }_3 }{72} \biggr)
 \la{finalZ2} \\ 
 & - & 
 \biggl( \frac{\gER^2 \Nc^{ }}{16\pi^2 } \biggr)^3
 \biggl( \frac{1097 \zeta^{ }_3 }{549} \biggr)
 \frac{61}{5}
 \biggl\{ 
     \ln \biggl( \frac{\bmu e^{\gammaE}}{4\pi T} \biggr)
 + 2 \ln\biggl( \frac{\bmu}{2\mER^{ }} \biggr)
 + \frac{\zeta'_3}{\zeta^{ }_3}
 - \gammaE 
 + \frac{103771}{52656}   
 \biggr\}
 \;, 
 \nn[3mm] 
 \delta Z^{ }_{\bg} & = & 
 - 
 \biggl( \frac{\gER^2 \Nc^{ }}{16\pi^2 } \biggr)^3
 \biggl( \frac{1097 \zeta^{ }_3 }{1098} \biggr)
 \frac{61}{5\epsilon}
 \;. \la{finaldZ2}
\ea
Remarkably, setting $\gER^2 = g^2\, (1 + \rmO(g^2))$, the divergence
in \eq\nr{finaldZ2} cancels 
1097/1098 of the coefficient of $1/\epsilon$  
in \eq\nr{ioan_eps}.
The remaining 1/1098 can be expressed as 
\be
 \delta \mathcal{Z}^{ }_{\bg} + \delta Z^{ }_{\bg}  
 \; = \;
 \frac{g^6 \Nc^3 T^2}{(8\pi)^2}
  \biggl( \frac{\zeta^{ }_3}{128\pi^4 T^2} \biggr)\, \frac{1}{45\epsilon} 
 + \rmO(g^8)
 \;, \la{finaldiv}
\ee
where in the round brackets we have isolated the master integral
in \eq\nr{Ib3}.

%
\subsection{Contribution from dimension-six operators in MQCD}
\la{ss:MQCD}

As already alluded to below \eq\nr{S_MQCD}, there are 
dimension-six operators also in MQCD. These originate from 
the purely spatial part of \eq\nr{chapman}, and also from 
1-loop effects within EQCD, as will be discussed 
in \se\ref{se:scalemE}.  
The corresponding action can be written as\footnote{%
 There are many alternative representations, for instance
 $
   \mathop{\rm tr} \bigl\{ F^{ }_{ij} F^{ }_{jk} F^{ }_{ki} \bigr\}
 = \frac{i \Nc^{ }}{2}f^{abc}_{ }F^{a}_{ij}F^{b}_{jk}F^c_{ki} 
 = \frac{i \Nc^{ }}{2(d-2)} f^{abc}_{ }\epsilon^{ }_{ijk}
   \widetilde{F}^a_i F^b_{jl} F^c_{kl}
 = \frac{i \Nc^{ }}{2} f^{abc}_{ }
   \widetilde{F}^a_i \widetilde{F}^b_j F^c_{ij}
 = \frac{i \Nc^{ }}{2(d-2)} f^{abc}_{ }\epsilon^{ }_{ijk}
 \widetilde{F}^a_i \widetilde{F}^b_j \widetilde{F}^c_k 
 $, 
 where we denoted the dual field strength by
 $ 
   \widetilde{F}^a_i \equiv \frac{\epsilon^{ }_{ijk}}{2} F^a_{jk}
 $
 and defined
 $ 
   \epsilon^{ }_{ijk} \epsilon^{ }_{lmn}
 \equiv \delta^{ }_{il} ( 
   \delta^{ }_{jm} \delta^{ }_{kn}
  - \delta^{ }_{jn} \delta^{ }_{km} ) 
 + \delta^{ }_{im} (
   \delta^{ }_{jn}\delta^{ }_{kl}
  - \delta^{ }_{jl}\delta^{ }_{kn} )
 + \delta^{ }_{in} (
   \delta^{ }_{jl} \delta^{ }_{km} 
  - \delta^{ }_{jm} \delta^{ }_{kl} 
  )
 $.
 } 
\ba
 \delta S^{ }_\rmii{MQCD}[A] & = & 
 2 \gM^2
 \int_X \mathop{\rm tr}
 \Bigl\{ 
  {\mathcal{C}}^{ }_1 \, (D^{ }_i F^{ }_{ij})^2 + 
 i\gM^{}  {\mathcal{C}}^{ }_3\, F^{ }_{ij} F^{ }_{jk} F^{ }_{ki}
 \Bigr\}
 \;, \hspace*{6mm} \la{chapman_MQCD}
\ea
where (recalling $\gM^2 = \gE^2\, (1 + \rmO(g))$) the hard contribution is 
$
 \delta \mathcal{C}^{ }_i \; = \; 
  \Tintip{P} {c^{ }_i} / {P^6}
$.

The dimension-six operators in \eq\nr{chapman_MQCD} give a contribution
to physical observables determined by MQCD, such as the spatial string
tension or ``magnetostatic'' screening masses. Given that MQCD
is a confining theory, these effects cannot be computed
analytically. We would like to know, however, whether the MQCD dynamics
can give an ultraviolet (UV) divergent contribution, compensating
against the term in \eq\nr{finaldiv}. 

In order to determine the UV divergence, we employ a trick similar to
that in ref.~\cite{aminusb}. All infrared (IR) contributions
are ``shielded'' by employing the propagators 
\be
 \langle A^a_k (p) A^b_l(q) \rangle 
 \;\equiv\; 
 \frac{\delta^{ab}\delta(p+q)}{p^2 + \mG^2}
 \biggl( 
 \delta^{ }_{kl} - \frac{\alpha\, p^{ }_k p^{ }_l }{p^2 + \mG^2}
 \biggr) \;, \quad
 \langle  c^a(p) {\bar c}^{\,b}(q)  \rangle
 \;\equiv\; 
  \frac{\delta^{ab}\delta(p-q)}{p^2 + \mG^2}
 \;, 
 \la{propags}
\ee
where $c^a, {\bar c}^{\,b}$ 
are ghost fields, $\alpha$ is a gauge parameter,  
and $\mG^{ } \equiv \gM^2 T/\pi$ is a fictitious mass. 
Once again, we compute a background field effective action, 
now denoted by $\Gamma^{ }_\rmii{IR}[B]$ given that 
the most IR fluctuations have been accounted for. 
We extract from it a 2-point function like in \eq\nr{G_MQCD_2}. 
The technical implementation follows that in 
\ses\ref{ss:1loop} and \ref{ss:2loop}.

Most contributions that we find are $\alpha$-dependent and void of physical 
significance. For instance, the 1-loop result has a structure similar to 
\eq\nr{MQCD_1l} but with $\mE^{ } \to \mG^{ }$:
\ba
 \left. \delta^{ } \Gamma^{(2)}_\rmii{IR} [B] 
 \right|^{ }_{\alpha = 0} & = & 
 \fr12 B^a_i(q)\, B^b_j(r)  \, \delta^{ab} \, \delta(q+r) \, 
 \gM^4 \Nc^2 \, I(\mG^{ })
 \nn & \times & 
 \biggl\{
  \bigl(  q^2 \delta^{ }_{ij} - q^{ }_i q^{ }_j  \bigr) \,
  \biggl[ 
    -\frac{11 \mathcal{C}^{ }_1}{3} + 18 \, \mathcal{C}^{ }_3
    + \rmO(\epsilon)
  \biggr] 
 \; + \; \rmO\Bigl(\frac{q^4}{\mG^2} \Bigr)
 \biggr\}
 \;. \la{1loopIR}
\ea 
This result is finite and proportional to 
$\mG^{ }$ and vanishes when we send $\mG^{ }\to 0$. 

However, at 2-loop order a non-trivial and gauge-independent result
emerges. Writing the contribution from Chapman 
vertices in a form reminiscent of \eq\nr{2loopgM5}, we get
\ba
 \delta^{ } \Gamma^{(2)}_\rmii{IR} [B] & = & 
 \fr12 B^a_i(q)\, B^b_j(r)  \, \delta^{ab} \, \delta(q+r) \, 
 \, 
 \gM^6 \Nc^3 \, H^{ }_{3}(\mG^{ })
 \nn & \times & 
 \biggl\{
  \frac{ \mG^2\, \delta^{ }_{ij}}{4d} \, D^{ }_1 
  \; + \; 
 \frac{ q^2 \delta^{ }_{ij} - q^{ }_i q^{ }_j }{4d} \, D^{ }_2 
  \; + \; 
 \frac{ q^{ }_i q^{ }_j }{4d} \, D^{ }_3 
 \; + \; \rmO\Bigl( \frac{q^4}{\mG^2} \Bigr)
 \biggr\}
 \;. \la{2loopIR}
\ea
The function $H^{ }_3$ is the three-mass variant of \eq\nr{H}, 
cf.\ \eq\nr{H3}, and has the same UV divergence, 
{\em viz.} $T^2 \mu^{-4\epsilon}/[(4\pi)^2 4\epsilon]$.
The coefficients $D^{ }_i$ contain a part 
$\propto H/H^{ }_3 = 1 +  \rmO(\epsilon)$.  
For $\epsilon \to 0$, $D^{ }_{1,3}$ are of $\rmO(\epsilon)$ 
and yield no divergence, 
whereas $D^{ }_2$ has a finite $\alpha$-independent part:  
\be
 D^{ }_2 = 24 \mathcal{C}^{ }_3 + \rmO(\epsilon)
 \;. \la{mqcd_res} 
\ee

Substituting 
$\mathcal{C}^{ }_3 \to  \Tintip{P} {c^{ }_3} / {P^6} $, 
inserting $c^{ }_3$ from \eq\nr{c1},    
and setting  $\gM^2 = g^2\mu^{2\epsilon}\, (1 + \rmO(g))$, 
yields a gauge-independent UV divergence and logarithmic part:
\ba
 \delta^{ } \Gamma^{(2)}_\rmii{IR} [B] & = &  
 \fr12 B^a_i (q)\, B^b_j (r)  \, \delta^{ab} \, \delta(q+r) \, 
 \bigl(  q^2 \delta^{ }_{ij} - q^{ }_i q^{ }_j  \bigr)
 \la{2loop_MQCD} \\ & \times & 
   \frac{ g^6 \Nc^3 T^2  }{(8\pi)^2} 
 \biggl( \frac{\zeta^{ }_3}{128\pi^4 T^2} \biggr)\, 
 \biggl( -\frac{1}{45}
 \biggr) 
 \biggl\{
  \frac{1}{\epsilon} 
  + 2 \ln \biggl( \frac{\bmu e^{\gammaE}}{4\pi T} \biggr)
  + 4 \ln \biggl( \frac{\bmu}{3\mG^{ }} \biggr)
  + \rmO(1) 
 \biggr\}
 \;.  \hspace*{5mm} \nonumber
\ea
Comparing with \eq\nr{finaldiv}, the divergence exactly cancels. 
Therefore we have now established our main technical goal, 
demonstrating that the 
IR-divergence in \eq\nr{ioan_eps} is fully cancelled by soft/hard
and ultrasoft/hard contributions from dimension-six operators. 

%
\section{Soft and overlapping ultrasoft/soft contributions}
\la{se:scalemE}

In \se\ref{se:scalepiT} we considered 
the soft/hard contributions to
the MQCD effective action, cf.\ \eq\nr{dZ2}. 
However, there are other contributions to 
$Z^{ }_{\bg}$, namely those associated with the purely ``soft'' 
contributions from the scale $\mE^{ }$. In order to distinguish
these from the effects considered in \se\ref{se:scalepiT}, we
denote them by $\widetilde{Z}^{ }_{\bg}$. 
For this section, we can take the super-renormalizable truncation 
in \eq\nr{S_EQCD} as a starting point, and $\mE^{ }$ as the only scale
being integrated out.

%
\subsection{Direct soft terms up to 3-loop level}

Up to 2-loop level, the value of 
$\widetilde{Z}^{ }_{\bg}$ was determined in ref.~\cite{pg2} 
(the dependence on scalar couplings 
was added in ref.~\cite{gE2}):\footnote{%
 In $d$ dimensions, 
 $
 \widetilde{Z}^{ }_{\bg} = 
 1 +  \gEs^2\Nc^{ }\! \int_p \frac{T}{6 (p^2 + \mEs^2)^2}
 + \gEs^4 \Nc^2
 \bigl[ \frac{d^3 - 10 d^2 + 23 d - 44}{3d(d-5)(d-7)} 
 - \frac{2\lambda}{3} \bigr]
 \int_p \frac{T}{p^2 + \mEs^2}
 \int_q \frac{T}{(q^2 + \mEs^2)^3}
 + \rmO( \gEs^6 \Nc^3 )
 $, 
 where the integrals are given in \eq\nr{masterI}.   
 } 
\be
 \widetilde{Z}^{ }_{\bg} \;=\; 
  1 \,+\,
    \frac{\gER^2 \Nc^{ } T}{48\pi \mER^{ }}
    \,+\, 
    \biggl( \frac{\gER^2 \Nc^{ } T}{16\pi \mER^{ }} \biggr)^2  
    \biggl( \frac{19}{18} + \frac{4\lambda}{3} \biggr)
         \,+\, \rmO\biggl( \frac{\gER^2 \Nc^{ } T}{16\pi \mER^{ }} \biggr)^3
 \;. \la{c2}
\ee
We now turn to the 3-loop contribution. 

The determination of 
$\widetilde{Z}^{ }_{\bg}$ is a rather straightforward exercise
in computer-algebraic methods for loop integrals. The Feynman
diagrams were generated with QGRAF~\cite{qgraf}. After expanding
in the external momentum and projecting onto the transverse and
longitudinal polarizations, we have to deal with vacuum-like 
master integrals. The 
subsequent simplifications, making use of renamings of 
integration variables and integration-by-parts (IBP) 
identities~\cite{ibp,ibp2}, 
have been programmed in FORM~\cite{form}. The values of 
the 3-loop master integrals can be found in refs.~\cite{aminusb,akr}
and are given in \eqs\nr{b_2} and \nr{b_4}. 
As a crosscheck, we have carried out two independent computations, 
whose results coincide perfectly. 
Our final ``bare'' expression reads\footnote{%
 The full $d$-dimensional form is given in appendix~E, 
 cf.\ \eqs\nr{fulld_first}--\nr{fulld_last}.
 }
\ba
 \delta^{ } \widetilde{\Gamma}^{(2)}_\rmii{MQCD} [B] & = &
 \fr12 B^a_i (q)\, B^b_j (r)  \, \delta^{ab} \, \delta(q+r) \, 
 \bigl(  q^2 \delta^{ }_{ij} - q^{ }_i q^{ }_j  \bigr)
 \biggl( \frac{\gE^2 \Nc^{ } T \mu^{-2\epsilon}}{16 \pi \mE^{ }} \biggr)^3
 \biggl( \frac{\bmu}{2 \mE^{ }} \biggr)^{6\epsilon}
 \nn 
 & \times &
 \biggl\{ 
  \frac{1 + 4 (\kappa_2^{ }- 4 \lambda)}{6\epsilon}
 \, + \, 
 \frac{2(23510 - 12600 \zeta^{ }_2 - 1101 \ln 2 )}{945}
 \nn 
 & + & 
 \frac{4\lambda + 24 \lambda^2
 - \kappa^{ }_1(5 - 8 \ln 2) 
 + \kappa^{ }_2(31 - 24 \ln 2)}{9} + \rmO(\epsilon)
 \biggr\} 
 \;. \la{gM2_3l_bare}
\ea

The $1/\epsilon$-divergences 
in \eq\nr{gM2_3l_bare} could {\em a priori} have 
an IR or UV origin. To find out, 
we have carried out the same computation by shielding all masses
like in \eq\nr{propags}, but with $\mG^{ }\to \mE^{ }$. Then only the 
divergence proportional to $4( \kappa_2^{ }- 4 \lambda )$ remains.  
This indicates that the divergence {\em not} containing scalar
self-couplings is purely of IR origin. 

We can envisage 
two possible sources for the IR divergence. One is related to 
ultrasoft contributions of the same type as in \se\ref{ss:MQCD}; these
are analyzed in \se\ref{se:MQCD}. The other is related to the mass 
parameter $\mE^{2}$.
It is well known that the physical Debye mass, defined as a screening 
mass related to a ``heavy-light'' state, is non-perturbative starting
at next-to-leading order~\cite{rebhan,ay}. Our $\mE^{2}$ is not 
such a physical mass but rather a Lagrangian parameter. 
Nevertheless, $\mE^2$ can still be considered IR sensitive 
at $\rmO(\gER^4 T^2)$. 
Indeed, if we compute the 2-point function
of $A^a_0$ at zero momentum, and shield all masses
like in \eq\nr{propags}, we find the UV divergence
cancelled by the mass counterterm in \eq\nr{mE_ct}. 
In contrast, if we compute the 2-point function without 
IR-shielding, we find an additional $1/\epsilon$-divergence
proportional to $\gER^4 T^2$, 
which depends on the gauge parameter $\alpha$.
This is an IR divergence,
i.e.\ $\sim \gER^4 T^2/\epsilon^{ }_\rmii{IR}$.

If we naively insert an ambiguity of this type
into the 1-loop term in \eq\nr{c2} and re-expand up 
to 3-loop order, the result is 
\be
 \frac{\gER^2 \Nc^{ } T }
 {48\pi \bigl[ \mER^{2} + \frac{\beta}{\epsilon^{ }_\rmiii{IR}}
 \bigl( \frac{\gER^2 \Nc^{ }T}{16 \pi } \bigr)^2
 \bigr]^{1/2}}
 \; -  \; 
 \frac{\gER^2 \Nc^{ } T}{48\pi \mER^{ }}
 \; \simeq \; 
 - \frac{\beta}{6 \epsilon^{ }_\rmiii{IR}}
 \biggl( \frac{\gER^2 \Nc^{ } T }{16\pi \mER} \biggr)^3
 \;. \la{beta}
\ee
On the non-perturbative level, $1/\epsilon^{ }_\rmii{IR}$ would turn
into a multiple of $\ln(c\, \mG^{ } / \mER^{ }) $, 
where 
$c$ is a non-perturbative constant
and 
the scale $\mG^{ }$ was defined around \eq\nr{propags}. 

Keeping in mind this expectation, 
we renormalize \eq\nr{gM2_3l_bare} 
by employing the proper mass counterterm from 
\eq\nr{mE_ct}. The UV divergences proportional to $\kappa^{ }_2 - 4\lambda$
duly cancel, and we find the 3-loop result
\ba
 \widetilde{Z}^{(3) }_{\bg} + 
 \delta \widetilde{Z}^{(3) }_{\bg} & = &
 \biggl( \frac{\gER^2 \Nc^{ } T }{16 \pi \mER^{ }} \biggr)^3
 \biggl\{ 
  \frac{1}{6\epsilon}
 + \biggl[ 1 + \frac{8(\kappa_2^{ }- 4 \lambda)}{3} \biggr]
   \ln\biggl( \frac{\bmu}{2 \mER^{ }} \biggr)
 \nn 
 & + & 
 \frac{2(23510 - 12600 \zeta^{ }_2 - 1101 \ln 2 )}{945}
 \nn 
 & + & 
 \frac{52\lambda + 24 \lambda^2
 - \kappa^{ }_1(5 - 8 \ln 2) 
 + \kappa^{ }_2(19 - 24 \ln 2)}{9}
 \biggr\} 
 \;. \la{gM2_3l}
\ea

%
\subsection{Contribution from dimension-six operators in MQCD}
\la{se:MQCD}

Parallelling \se\ref{ss:MQCD}, let us finally consider contributions
from ultrasoft effects to the gauge coupling, 
in the presence of 
dimension-six operators in MQCD. The action has the form in 
\eq\nr{chapman_MQCD}, with the coefficients now completed to include
the soft contribution:  
\be
 \mathcal{C}^{ }_i \; = \; 
  \Tint{P}' \frac{c^{ }_i}{P^6}
 + T\int_p \frac{\tilde{c}^{ }_i}{(p^2 + \mE^2)^3} 
 \;, \quad i=1,3
 \;. \la{Ci}
\ee
The spatial integral appearing is related to that in \eq\nr{I}
as shown by \eq\nr{masterI}, 
\be
 \int_{p} \frac{T}{(p^2 + \mE^2)^3} 
 \; = \; 
 \frac{\mE^{d-6}\Gamma(3-\frac{d}{2})T}{2 (4\pi)^{\frac{d}{2}}}
 \; \stackrel{3-2\epsilon}{=} \;
 \frac{T \mu^{-2\epsilon}_{ }}{32\pi\mE^3}\, 
 \biggl[ 1 + 2\epsilon\,\Bigl( 1 + \ln\frac{\bmu}{2 \mE^{ }}\Bigr)  
 + \rmO(\epsilon^2) \biggr]
 \;. \la{I3}
\ee

Including the overall prefactor from \eq\nr{chapman_MQCD} and
the integral from \eq\nr{I3}, 
the new contributions to the coefficients of the dimension-six operators
are $\sim \gM^2 T/ \mE^3$ at 1-loop level. 
Including a fictitious IR-regulator like in \eq\nr{propags}, 
the 1-loop contribution from these operators 
to $\widetilde{Z}^{ }_{\bg}$ 
comes with a factor $\sim \gM^2 T \mG^{ }$ and vanishes
for $\mG^{ }\to 0$, whereas the 2-loop contribution comes with a factor
$\sim \gM^4 T^2$ and can yield a contribution
$\sim \gM^6 T^3/ \mE^3 \sim \rmO(g^3)$ to~$\widetilde{Z}^{ }_{\bg}$.
2-loop contributions to the coefficients of dimension-six operators
would be $\sim \gM^4 T^2/ \mE^4$ and therefore lead to effects 
suppressed by $\sim \rmO(g^4)$. 
Dimension-eight operators, 
whose coefficients are $\sim \gM^2 T/ \mE^5$, 
lead to effects suppressed by 
$\sim \gM^{10} T^5/ \mE^5 \sim \rmO(g^5)$.

%
\begin{figure}[t]
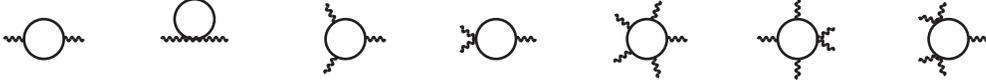


\hspace*{0.2cm}%
\begin{minipage}[c]{14cm}
\begin{eqnarray*}
 && 
 \DiagBma \quad
 \DiagBmb \quad
 \DiagBmc \quad
 \DiagBmd \quad
 \DiagBsc \quad
 \DiagDsc \quad
 \DiagFsc 
\end{eqnarray*}
\end{minipage}

\caption[a]{\small 
1-loop contributions to the MQCD 2-point, 3-point and 5-point 
functions in the background field gauge.
Wiggly lines denote ultrasoft gluons 
and solid lines adjoint scalars. 
}
\la{fig:5pt_EQCD}
\end{figure}
%
 
According to \eq\nr{3pt}, the value of 
$\tilde c^{ }_1$ can be inferred from the 2-point and that of 
$\tilde c^{ }_3$ from the 3-point vertex of the background
field effective action. 
To be sure that no operators got overlooked, 
we have also determined them from the 5-point vertex, cf.\ the 
spatial part of  \eq\nr{5pt},  
which leads to several independent crosschecks
(the diagrams are shown in \fig\ref{fig:5pt_EQCD}). 
We find that the results are related in a curious way to the $d$-dependence
of $c^{ }_1$ and $c^{ }_3$ in \eq\nr{c1}:\footnote{%
 To our knowledge
 these values were first obtained for $d=3$
 by P.~Giovannangeli (unpublished, 2005), along lines  
 that have recently been documented in ref.~\cite{cpa}. 
 }${}^{,}$\footnote{%
 We note in passing that 
 even though the $\tilde{c}^{ }_i$ contribution in \eq\nr{Ci}
 is parametrically larger  
 by $\rmO(1/g^3)$ than the $c^{ }_i$ contribution, 
 the large value of $c^{ }_1$ in \eq\nr{c1} implies that
 numerically $c^{ }_1$ and $\tilde{c}^{ }_1$ give similar contributions  
 if $g^2 \sim 2$. If $g^2 \gg 1$, 
 $\mathcal{C}^{ }_1$ becomes positive. 
 }
\be
 \tilde c^{ }_1 \; = \; -\frac{1}{120} \;, \quad
 \tilde c^{ }_3\;  = \; -\frac{1}{180} \;.  
 \la{tc1}
\ee 

Inserting these values into \eq\nr{mqcd_res}, and
substituting $\gM^2 = \gER^2\mu^{2\epsilon}\, (1 + \rmO(g))$, 
we find 
a gauge-independent UV divergence and logarithmic part:
\ba
 \delta^{ } \widetilde{\Gamma}^{(2)}_\rmii{IR} [B]
 & = &  
 \fr12 B^a_i (q)\, B^b_j (r)  \, \delta^{ab} \, \delta(q+r) \, 
 \bigl(  q^2 \delta^{ }_{ij} - q^{ }_i q^{ }_j  \bigr)
 \nn
 & \times &   
 \biggl( \frac{ \gER^2 \Nc^{ } T }{16 \pi\mER^{ }} \biggr)^3
 \biggl( 
  - 
  \frac{1}{45}
 \biggr) 
 \biggl\{ 
   \frac{1}{\epsilon} 
  + 2 \ln \biggl( \frac{\bmu}{2 \mER^{ }} \biggr)
  + 4 \ln \biggl( \frac{\bmu}{3\mG^{ }} \biggr)
  + \rmO(1)  
 \biggr\}
 \;.  \hspace*{5mm} \la{2loop_MQCD_tilde}
\ea
This implies that the counterterm needed in MQCD reads
$
 \delta \widetilde{Z}^{(3) }_{\bg} = 
 \bigl( \frac{ \gER^2 \Nc^{ } T }{16 \pi\mER^{ }} \bigr)^3
  \frac{1}{45\epsilon}
$.

Obviously, 
\eq\nr{2loop_MQCD_tilde}
does {\em not} match
the divergence in \eq\nr{gM2_3l}. 
In other words, if we subtract the part needed to serve as
$ \delta \widetilde{Z}^{(3) }_{\bg} $ from \eq\nr{gM2_3l},
an IR divergence remains.
In terms of the coefficient $\beta$ introduced in~\eq\nr{beta}, 
it amounts to  $\beta = -\frac{13}{15}$. Let us stress 
that we have verified the gauge independence of this result. 
Therefore we are left to speculate that a non-perturbative
mass ambiguity of the type discussed around \eq\nr{beta} prohibits
a purely perturbative determination of 
$
 \widetilde{Z}^{(3) }_{\bg} 
$, 
and thus of $\gM^2$
in terms of $\gER^2$ and $\mER^{ }$
at $\rmO(\gER^6 T^3/ \mER^3)$.

%
\section{Conclusions} 
\la{se:concl}

The main technical ingredient of this investigation 
was the analysis carried out in \se\ref{se:scalepiT}. We considered
dimension-six operators induced by integrating out the ``hard'' 
momenta $\sim\pi T$ from thermal QCD~\cite{chapman}. Specifically, 
we computed at 1-loop and 2-loop levels the influence of 
these operators on the  gauge coupling felt by ultrasoft 
(magnetostatic) modes. 
Remarkably, including UV divergences originating 
both from ``soft'' loops at the Debye 
scale $\mE^{ }\sim gT$ and ``ultrasoft'' loops at the 
non-perturbative scale $\sim g^2 T/\pi$, we observed an exact
cancellation of the IR divergence found in a 3-loop
determination of the EQCD gauge coupling
(cf.\ \eq\nr{ioan_eps})~\cite{ig,ig_gE}. This 
represents a nice crosscheck 
of the effective theory setup as a whole. 

As a second technical ingredient, discussed in \se\ref{se:scalemE}, we 
considered the ``soft'' contributions to the ultrasoft gauge coupling. 
We determined direct 3-loop effects
(cf.\ \eq\nr{gM2_3l}) and compared them with overlapping ultrasoft/soft
contributions originating from dimension-six operators induced by
integrating out the soft momenta $\sim \mE^{ }$
(cf.\ \eq\nr{2loop_MQCD_tilde}). This time only a partial
cancellation of soft IR divergences against ultrasoft/soft UV divergences
was observed. As a culprit, we speculate
that a non-perturbative ambiguity of the soft scale within 
EQCD sets an upper bound on the accuracy with which effects 
depending on $\mE^{ }$ can be determined within perturbation theory. 
This may be surprising insofar as no such problem was met in 3-loop or  
4-loop studies of the EQCD vacuum energy density~\cite{bn,aminusb}. 
However, the present quantity is different, being not directly 
a physical observable but rather an effective Lagrangian parameter
(the MQCD gauge coupling $\gM^2$). 

On a more general level, the main conclusions that we draw are as follows: 
\bi

\item[(i)] 
Even if the colour-electric 
scale $\mE^{ } \sim g T$ is formally larger 
than the colour-magnetic scale $\sim g^2 T/\pi$, it does 
play an essential role in the IR dynamics. Concretely, 
in terms of the IR divergence found by integrating out the 
hard scale $\sim \pi T$, the colour-electric scale 
is 1097 times more important than the colour-magnetic scale
(cf.\ \eq\nr{finaldZ2}). 

\item[(ii)] 
Dimension-six operators need to be included in 
EQCD if good precision is required. Indeed, as we
have demonstrated analytically (cf.\ point~(i)),  
they do influence the IR dynamics of the system. This is a possible
reason for why the super-renormalizable truncation of EQCD 
fails close to~$\Tc^{ }$ even in pure Yang-Mills 
theory~\cite{lattg7}. 

\item[(iii)] 
Apart from the indications in point~(i) that the scale $\mE^{ }$
is important, we also find trouble if we try to integrate 
it out. The reason could be that EQCD is a confining theory, and that
physics at the scale $\mE^{2}$ should in general be 
affected by non-perturbative ambiguities of $\rmO(g^4T^2/\pi^2)$.
Once $\mE^{ }$ is integrated out, some remnant of these
ambiguities may remain, if the parameters of MQCD are determined
up to the corresponding relative precision. It would be interesting
to find a way to determine the leading non-perturbative
contribution to $\gM^2$ through lattice methods, even if this
requires the simultaneous inclusion of 
the $1/\mE^3$-suppressed MQCD dimension-six operators 
in \eq\nr{chapman_MQCD}. 

\ei

%
\section*{Acknowledgements}

This work was partly supported by the Swiss National Science Foundation
(SNF) under grant 200020-168988, by the FONDECYT under project 1151281, 
and by the UBB under project GI-172309/C.

%
\appendix
\renewcommand{\thesection}{Appendix~\Alph{section}}
\renewcommand{\thesubsection}{\Alph{section}.\arabic{subsection}}
\renewcommand{\theequation}{\Alph{section}.\arabic{equation}}

%
\section{Spacetime and colour tensors}

Because the presence of a heat bath breaks Lorentz invariance, 
we need to introduce separate notation for spatial and zero 
spacetime indices. The full Kronecker symbol is denoted by
\be
 \delta^{ }_{\mu\nu} \; \equiv \; T^{ }_{\mu\nu} + S^{ }_{\mu\nu}
 \;,
 \quad 
 T^{ }_{\mu\nu} \; \equiv \; \delta^{ }_{\mu 0}\delta^{ }_{\nu 0} 
 \;,
 \quad 
 S^{ }_{\mu\nu} \; \equiv \;  \delta^{ }_{\mu i}\delta^{ }_{\nu i}
 \;. \la{Smunu}
\ee
We also introduce the totally symmetric tensors
\ba
 T^{ }_{\mu\nu\rho\sigma} & \equiv &  
\delta^{ }_{\mu 0} \delta^{ }_{\nu 0}
\delta^{ }_{\rho 0} \delta^{ }_{\sigma 0}
 \;, \\ 
 T^{ }_{\mu\nu\rho\sigma\alpha\beta} & \equiv &  
 \delta^{ }_{\mu 0} \delta^{ }_{\nu 0}
 \delta^{ }_{\rho 0} \delta^{ }_{\sigma 0}
 \delta^{ }_{\alpha 0} \delta^{ }_{\beta 0}
 \;, \\ 
 \delta^{ }_{\mu\nu\rho\sigma} & \equiv &  
 \delta^{ }_{\mu \nu} \delta^{ }_{\rho \sigma}  
 + \mbox{2 permutations} 
 \;, \\ 
 \delta^{ }_{\mu\nu\rho\sigma\alpha\beta} & \equiv &  
 \delta^{ }_{\mu \nu}  
 \delta^{ }_{\rho \sigma}  
 \delta^{ }_{\alpha \beta}  
 + \mbox{14 permutations} 
 \;. \la{perm14}
\ea

For the colour indices, it is helpful to denote
\be
 X_{ }^{a_1 a_2 ... a_n} \; \equiv \; 
 f_{ }^{m_n a_1 m_1 }
 f_{ }^{m_1 a_2 m_2 }
  \cdots
 f_{ }^{m_{n-1} a_n m_n }
 \;, \la{def_X}
\ee
as well as the symmetrized versions 
\be
  X_{ }^{\{a_1 ...  a_2\} ... } \; \equiv \; 
 \fr12 \bigl( X_{ }^{a_1 ... a_2 ... } + X_{ }^{a_2 ... a_1 ... }\bigr)
 \;, \quad
  X_{ }^{[a_1 ... a_2] ... } \; \equiv \; 
 \fr12 \bigl( X_{ }^{a_1 ... a_2 ... } - X_{ }^{a_2 ... a_1 ... }\bigr)
 \;. 
\ee
These objects satisfy 
$
 X_{ }^{a_n a_{n-1}... a_2\, a_1} = (-1)^n  X_{ }^{a_1 a_2 ... a_{n-1} a_n} 
$,
$
 X_{ }^{a_1 a_2 ... a_{n-1} a_n} = X_{ }^{a_2 ... a_{n-1} a_n a_1}
$.
It follows that 
\be
 X_{ }^{\{a_1 a_2\}[ a_3 a_4 ]} =
 X_{ }^{\{a_1 a_2\}\{ a_3 a_4 a_5\} } =
 X_{ }^{[a_1 a_2][ a_3 a_4 a_5] } =
 X_{ }^{\{a_1 a_2 a_3\}[ a_4 a_5 a_6 ]} =
  0
 \;.  
\ee
Therefore we can write
\ba
 X_{ }^{a_1 a_2 a_3 a_4} & = & 
  X_{ }^{\{a_1 a_2\}\{ a_3 a_4\} } + 
  X_{ }^{[a_1 a_2][ a_3 a_4 ]}
 \;,  \\
 X_{ }^{a_1 a_2 a_3 a_4 a_5} & = & 
 X_{ }^{\{a_1 a_2\}[ a_3 a_4 a_5] } +
 X_{ }^{[a_1 a_2]\{ a_3 a_4 a_5\} } 
 \;,  \\
 X_{ }^{a_1 a_2 a_3 a_4 a_5 a_6} & = & 
 X_{ }^{\{a_1 a_2 a_3\}\{ a_4 a_5 a_6 \} } +
 X_{ }^{[a_1 a_2 a_3] [ a_4 a_5 a_6] } 
 \;.
\ea
It may furthermore be noted that
\ba
 && X_{ }^{a_1 a_2 a_3} =   
 - \frac{\Nc^{ }}{2} f_{ }^{a_1 a_2 a_3}
 \;, \quad
 X_{ }^{[a_1 a_2][ a_3 a_4 ]} = 
 - \frac{\Nc^{ }}{4}  f_{ }^{ m a_1 a_2} f_{ }^{m a_3 a_4}  
 \;, \\
 && X_{ }^{[a_1 a_2]a_3[ a_4 a_5 ]}  =   
 - \frac{\Nc^{ }}{8} f_{ }^{ m a_1 a_2} f_{ }^{m a_3 n} f_{ }^{n a_4 a_5} 
 \;, \\[1mm] 
 && f^{a_1 a_2 n}_{ } X^{n a_3 a_4 ... }_{ } =
 2 X_{ }^{[a_1 a_2] a_3 a_4 ... } 
 = 
 X_{ }^{a_1 a_2 a_3 a_4 ... } - 
 X_{ }^{a_2 a_1 a_3 a_4 ... }
 \;. 
\ea

%
\section{Basic sum-integrals}

Employing the notation defined in \eqs\nr{Smunu}--\nr{perm14}, 
the following relations can be established:
\ba
 \Tint{P}' \frac{P_\mu P_\nu}{P^4} & = &
 \Tint{P}' \frac{(1-d)\, T^{ }_{\mu \nu} + 
 \delta^{ }_{\mu \nu} }{2 P^2} 
 \;, \\  
 \Tint{P}' \frac{P_\mu P_\nu}{P^6} & = &
 \Tint{P}' \frac{(3-d)\, T^{ }_{\mu \nu} + 
 \delta^{ }_{\mu \nu} }{4 P^4} 
 \;, \\ 
 \Tint{P}' \frac{P_\mu P_\nu}{P^8} & = &
 \Tint{P}' \frac{(5-d)\, T^{ }_{\mu \nu} + 
 \delta^{ }_{\mu \nu} }{6 P^6} 
 \;, \\
 \Tint{P}' \frac{P_\mu P_\nu P_\rho P_\sigma}{P^8} & = &
 \Tint{P}' \biggl\{ \frac{(3-d)(1-d) 
 T^{ }_{\mu \nu \rho \sigma}
 }{24P^4}
 \nn &  & \hspace*{-2cm} + \, \frac{ 
 (3-d)\,
  ( T^{ }_{\mu \nu} \delta^{ }_{\rho \sigma} 
 + \mbox{5 permutations}
 )
  + 
 \delta^{ }_{\mu\nu\rho\sigma}
 }{24 P^4} \biggr\}
 \;, \hspace*{5mm} \\ 
 \Tint{P}' \frac{P_\mu P_\nu P_\rho P_\sigma}{P^{10}} & = &
 \Tint{P}' \biggl\{
  \frac{(5-d)(3-d)\,  T^{ }_{\mu\nu\rho\sigma}
  }{48 P^6} 
 \nn &  & \hspace*{-2cm} + \, \frac{ 
 (5-d)\,
  ( T^{ }_{\mu \nu} \delta^{ }_{\rho \sigma} 
 + \mbox{5 permutations}
 )
  + 
  \delta^{ }_{\mu\nu\rho\sigma}
 }{48 P^6} \biggr\}
 \;, \hspace*{5mm} \\
 \Tint{P}' \frac{P_\mu P_\nu P_\rho P_\sigma P_\alpha P_\beta}{P^{12}} & = &
 \Tint{P}' \biggl\{ 
  \frac{(5-d)(3-d)(1-d)\,  T^{ }_{\mu\nu\rho\sigma\alpha\beta}
  }{480 P^6} \nn 
 & & \hspace*{-2cm} + \, 
  \frac{(5-d)(3-d)\, 
 ( T^{ }_{\mu \nu \rho \sigma} \delta^{ }_{\alpha\beta}
 + \mbox{14 permutations}
  )
 }{480 P^6}  \nn 
 & & \hspace*{-2cm} + \, 
  \frac{(5-d)\, 
 ( T^{ }_{\mu \nu} \delta^{ }_{\rho\sigma\alpha\beta}
 + \mbox{14 permutations}
  )
  + 
  \delta^{ }_{\mu\nu\rho\sigma\alpha\beta}
 }{480 P^6} \biggr\}
 \;. \hspace*{5mm} 
\ea
These are needed for the computations in \se\ref{ss:determine}. 

%
\section{Dimension-six vertices in the $S/T$ basis}
\la{ss:v2}

In \se\ref{ss:determine} we displayed (parts of) 
the vertices originating from \eq\nr{chapman} in a basis in which
spacetime indices appear in the form similar to appendix~B. 
For the considerations of \se\ref{se:scalepiT}, it is 
advantageous to employ a basis in which the spatial and temporal
indices are strictly separated from each other. This can be 
implemented with the tensors 
$S^{ }_{\mu\nu\cdots}$ and $T^{ }_{\mu\nu\cdots}$, 
defined in \eq\nr{Smunu}. In this section we display all 
the Chapman vertices originating from \eq\nr{chapman} with such a notation. 

The 2-point Chapman vertex reads 
\ba
 \delta S^{(2)}_\rmii{EQCD}  & = & 
 A^a_\mu(q)\, A^a_\nu(-q)  \, 
 \biggl(  \Tint{P}'\frac{\gE^2 \Nc^{ }}{P^6} \biggr)
 \Bigl\{
  \varzeta^{ }_1 \, q^2 \bigl( q^2 S^{ }_{\mu\nu} - q^{ }_\mu q^{ }_\nu \bigr) 
  + \varzeta^{ }_2 \, q^4 T^{ }_{\mu\nu}
 \Bigr\} 
 \;, \la{2pt_ops}
\ea
where
\be
 \varzeta^{ }_{1} = 2 c^{ }_1 \;, \quad
 \varzeta^{ }_{2} = 2(c^{ }_1 + c^{ }_2)
 \;. \la{sc_2}
\ee
The 3-point Chapman vertex becomes  
\ba
 \delta S^{(3)}_\rmii{EQCD} & = & 
  A^a_\mu(q)\, A^b_\nu(r)\, A^c_\rho(s)
 \, f^{abc} \, \delta(q+r+s)  
 \biggl(  \Tint{P}'\frac{ i \gE^3 \Nc^{ } }{P^6} \biggr)
 \nn & \times & \biggl\{ 
  \xi^{ }_1\, q^{ }_\mu q^{ }_\nu q^{ }_\rho
 + \xi^{ }_2 \, q^{ }_\mu q^{ }_\nu r^{ }_\rho
 + \xi^{ }_3\, q^{ }_\mu r^{ }_\nu q^{ }_\rho
 + \xi^{ }_4 \, r^{ }_\mu q^{ }_\nu q^{ }_\rho 
 \nn & + & 
 S^{ }_{\mu\nu} \Bigl[ \xi^{ }_5 \, q^2 q^{ }_\rho 
                     + \xi^{ }_6 \, q^2 r^{ }_\rho
                     + \xi^{ }_7 \, s^2 q^{ }_\rho \Bigr] 
 + 
 T^{ }_{\mu\nu} \Bigl[ \xi^{ }_8 \, q^2 q^{ }_\rho
                      + \xi^{ }_9 \, q^2 r^{ }_\rho
                     + \xi^{ }_{10} \, s^2 q^{ }_\rho \Bigr] 
 \biggr\}
 \;, \hspace*{9mm} \la{3pt_ops}
\ea
where $q^{ }_\mu q^{ }_\nu q^{ }_\rho$ and
$ q^{ }_\mu q^{ }_\nu q^{ }_\rho + 
  q^{ }_\mu q^{ }_\nu r^{ }_\rho -
  q^{ }_\mu r^{ }_\nu q^{ }_\rho  = 
  - q^{ }_\mu ( q^{ }_\nu s^{ }_\rho + r^{ }_\nu q^{ }_\rho ) $
actually vanish as can be seen by the relabelling 
$ (r\leftrightarrow s, \nu \leftrightarrow \rho, b\leftrightarrow c)$.
Therefore any change $\delta \xi^{ }_1$ or any simultaneous 
change $\delta \xi^{ }_2  = - \delta \xi^{ }_3$ has no effect.
It can be checked that \eqs\nr{MQCD_1l} and 
\nr{C1}--\nr{C3} are invariant in these transformations.
A representation of the coefficients can be chosen as 
\ba
 && \xi^{ }_1 = 0 \;,\quad
    \xi^{ }_2 = 2 c^{ }_3 \;,\quad
    \xi^{ }_3 = - 4 c^{ }_1 \;,\quad
    \xi^{ }_4 = - 2 c^{ }_3 \;,
 \nn 
 && \xi^{ }_5 = -3 c^{ }_3 \;,\quad
    \xi^{ }_6 = 8 c^{ }_1 -3 c^{ }_3 \;,\quad
    \xi^{ }_7 = 3 c^{ }_3 - 4 c^{ }_1 \;,\quad
    \xi^{ }_8 = -4c^{ }_2 -3 c^{ }_3 - c^{ }_4 + c^{ }_5 \;,
 \nn 
 &&   \xi^{ }_9 = 8 c^{ }_1 + 4c^{ }_2  -3 c^{ }_3 - c^{ }_4 + c^{ }_5 \;,\quad
    \xi^{ }_{10} = 3 c^{ }_3 - 4 c^{ }_1 + c^{ }_4 - c^{ }_5
 \;. \la{sc_3}
\ea

The 4-point vertex amounts to 
\ba
 \delta S^{(4)}_\rmii{EQCD}  & = & 
  A^a_\mu(q)\, A^b_\nu(r)\, A^c_\alpha(s)\, A^d_\beta(t)
  \, \delta(q+r+s+t)  
 \biggl(  \Tint{P}'\frac{ \gE^4 }{P^6} \biggr)
 \nn & \times & \biggl\{ 
 X^{\{ab\}\{cd\}}\, 
 \bigl[
  S^{ }_{\mu\alpha}S^{ }_{\nu\beta} \, 
  \bigl( \psi^{ }_1\, q^2 + \psi^{ }_3\, q\cdot r\bigr)   
 \nn & & \hspace*{1.7cm} + \, 
  T^{ }_{\mu\alpha}S^{ }_{\nu\beta} \, 
  \bigl( \psi^{ }_4\, q^2 + \psi^{ }_5 \, r^2 + \psi^{ }_6\, q\cdot r\bigr)   
 \nn & & \hspace*{1.7cm} + \, 
  S^{ }_{\mu\nu}S^{ }_{\alpha\beta} \, 
  \bigl( \psi^{ }_{10}\, q^2 + \psi^{ }_{12}\, q\cdot r\bigr)   
  \; + \; 
  T^{ }_{\mu\nu}S^{ }_{\alpha\beta} \, 
  \bigl( \psi^{ }_{13}\, q^2 + \psi^{ }_{15}\, q\cdot r\bigr)   
 \nn & & \hspace*{1.7cm} + \, 
  S^{ }_{\mu\nu}T^{ }_{\alpha\beta} \, 
  \bigl( \psi^{ }_{16}\, q^2 + \psi^{ }_{18}\, q\cdot r\bigr)   
  \; + \; 
  T^{ }_{\mu\nu \alpha\beta} \, 
  \bigl( \psi^{ }_{19}\, q^2 + \psi^{ }_{21}\, q\cdot r\bigr)   
 \nn & & \hspace*{1.7cm} + \, 
  S^{ }_{\mu\alpha} \, 
  \bigl( \psi^{ }_{22}\, q^{ }_\nu q^{ }_\beta
   + \psi^{ }_{23} \, q^{ }_\nu r^{ }_\beta
       + \psi^{ }_{24}\, r^{ }_\nu q^{ }_\beta
 + \psi^{ }_{25} \, r^{ }_\nu r^{ }_\beta \bigr)   
 \nn & & \hspace*{1.7cm} + \, 
  T^{ }_{\mu\alpha} \, 
  \bigl( \psi^{ }_{26}\, q^{ }_\nu q^{ }_\beta
  + \psi^{ }_{27} \, q^{ }_\nu r^{ }_\beta
       + \psi^{ }_{28}\, r^{ }_\nu q^{ }_\beta
 + \psi^{ }_{29} \, r^{ }_\nu r^{ }_\beta \bigr)   
 \nn & & \hspace*{1.7cm} + \, 
  S^{ }_{\mu\nu} \, 
  \bigl( \psi^{ }_{30}\, q^{ }_\alpha q^{ }_\beta
  + \psi^{ }_{31} \, q^{ }_\alpha r^{ }_\beta
    \bigr)   
  \; + \; 
  T^{ }_{\mu\nu} \, 
  \bigl( \psi^{ }_{34}\, q^{ }_\alpha q^{ }_\beta
  + \psi^{ }_{35} \, q^{ }_\alpha r^{ }_\beta
    \bigr)   
 \nn & & \hspace*{1.7cm} + \, 
  S^{ }_{\alpha\beta} \, 
  \bigl( \psi^{ }_{38}\, q^{ }_\mu q^{ }_\nu
      + \psi^{ }_{39} \, q^{ }_\mu r^{ }_\nu
       + \psi^{ }_{40}\, r^{ }_\mu q^{ }_\nu 
  \bigr)   
 \nn & & \hspace*{1.7cm} + \, 
  T^{ }_{\alpha\beta} \, 
  \bigl( \psi^{ }_{42}\, q^{ }_\mu q^{ }_\nu
   + \psi^{ }_{43} \, q^{ }_\mu r^{ }_\nu
       + \psi^{ }_{44}\, r^{ }_\mu q^{ }_\nu
  \bigr)   
 \bigr] 
 \nn 
 & + & \hspace*{3mm}
 X^{[ab][cd]}\, 
 \bigl[ \psi^{ }_i \to \omega^{ }_i 
 \bigr]
 \biggr\}
 \;,  \la{4pt_ops}
\ea
where some coefficients have been dropped because they can be 
converted to the remaining ones through trivial renamings
of indices and integration variables. 
The values are
\ba
 && \psi^{ }_1 =  0 \;, \quad \psi^{ }_3 = -8c^{ }_1\;, \nn
 && \psi^{ }_4 =  0 \;, \quad \psi^{ }_5 = 0 
                \;, \quad \psi^{ }_6 = -16c^{ }_1-4 c^{ }_5+8c^{ }_7\;, \nn
 &&   \psi^{ }_{10}=-4c^{ }_1 \;, \quad \psi^{ }_{12} = -4c^{ }_1\;,
  \nn && 
   \psi^{ }_{13}=  0  \;, \quad
   \psi^{ }_{15} = 0 \;, \quad
  \psi^{ }_{16}=-8c^{ }_1-2c^{ }_5 +4c^{ }_7 
  \;, \quad 
  \psi^{ }_{18} = -8c^{ }_1 - 2c^{ }_5 - 4c^{ }_6\;, \nn
 && \psi^{ }_{19}=-4c^{ }_1-2c^{ }_5+ 4 c^{ }_7  +2c^{ }_9 
    \;, \quad \psi^{ }_{21} = -12c^{ }_1-6c^{ }_5
  -4c^{ }_6+8c^{ }_7-2c^{ }_8+4c^{ }_9\;, \nn
 && \psi^{ }_{22} = -8c^{ }_1 \;, \quad \psi^{ }_{23} = 12c^{ }_1\;, \quad 
     \psi^{ }_{24} = -4c^{ }_1\;,  \quad \psi^{ }_{25} = 4c^{ }_1 \;, \nn  
 && \psi^{ }_{26} = -8c^{ }_1-8c^{ }_2 \;, \quad 
    \psi^{ }_{27} = 12c^{ }_1-20c^{ }_2+8c^{ }_5-16c^{ }_7\;, \nn 
 &&  \psi^{ }_{28} = -4c^{ }_1 + 12 c^{ }_2 -  4c^{ }_5+8 c^{ }_7\;,  \quad 
     \psi^{ }_{29} = 4c^{ }_1 +4c^{ }_2 \;, \nn  
 && \psi^{ }_{30} = 4c^{ }_1 \;, \quad 
    \psi^{ }_{31} = -4c^{ }_1 \;,\quad  
    \psi^{ }_{34} = 4c^{ }_1 + 4c^{ }_2 \;, \quad 
    \psi^{ }_{35} = -4c^{ }_1-4 c^{ }_2 \;,  \nn  
 && \psi^{ }_{38} = 4 c^{ }_1 \;, \quad \psi^{ }_{39} = 0 \;, \quad 
     \psi^{ }_{40} = 8 c^{ }_1 \;,  \nn  
 && \psi^{ }_{42} = 4 c^{ }_1 - 4 c^{ }_2 + 2 c^{ }_5 - 4 c^{ }_7 \;, \quad 
    \psi^{ }_{43} = 8 c^{ }_2 - 2 c^{ }_5 +4 c^{ }_7\;, \nn
 && 
    \psi^{ }_{44} = 8 c^{ }_1 - 8 c^{ }_2
  + 4 c^{ }_5 +4 c^{ }_6 - 4c^{ }_7\;,   \nn  
 && \omega^{ }_1 =  -16 c^{ }_1 \;,
 \quad \omega^{ }_3 = 8 c^{ }_1 - 12 c^{ }_3 \;, \nn
 && \omega^{ }_4 =  -16 c^{ }_1 - 16 c^{ }_2 \;, \quad 
    \omega^{ }_5 = -16 c^{ }_1 -4c^{ }_5 + 8 c^{ }_7  \;, \nn
 && 
    \omega^{ }_6 = 16 c^{ }_1 - 24 c^{ }_3 - 8 c^{ }_4
  + 4 c^{ }_5 + 8 c^{ }_7 \;, \nn
 && \omega^{ }_{22} = -24 c^{ }_1 \;, \quad
   \omega^{ }_{23} = -44 c^{ }_1 + 24 c^{ }_3\;, \quad 
     \omega^{ }_{24} = -12 c^{ }_1 \;,
  \quad \omega^{ }_{25} = 4 c^{ }_1 \;, \nn  
 && \omega^{ }_{26} = -24 c^{ }_1 - 24 c^{ }_2 
  \;, \quad \omega^{ }_{27} = -44 c^{ }_1 -12 c^{ }_2
  + 24 c^{ }_3 + 8 c^{ }_4 -8c^{ }_5\;, \nn
 &&   \omega^{ }_{28} = -12 c^{ }_1 - 28 c^{ }_2 + 4 c^{ }_5 -8c^{ }_7 
 \;,  \quad \omega^{ }_{29} = 4 c^{ }_1 - 12 c^{ }_2
  + 4 c^{ }_5 - 8 c^{ }_7 \;, \nn  
 && \omega^{ }_{30} =  0
 \;, \quad \omega^{ }_{31} = 20 c^{ }_1 - 12 c^{ }_3\;, \quad 
    \omega^{ }_{34} =  0 \;, \nn 
 &&   \omega^{ }_{35} = 20 c^{ }_1 + 20 c^{ }_2
  - 12 c^{ }_3 - 4 c^{ }_4  + 8 c^{ }_7
 \;. \la{sc_4}
\ea
In the case of $\omega^{ }_i$, all coefficients associated with operators
containing $S^{ }_{\alpha\beta}$ or $T^{ }_{\alpha\beta}$ vanish, because
of antisymmetry. 

The coefficients of the 4-point vertex 
listed above are {\em not} independent. Indeed 
momentum conservation leads to relations between the different structures
defined in \eq\nr{4pt_ops}, which 
implies that certain linear combinations of the coefficients couple to null 
operators. In the spirit of \eq\nr{Theta_c}, 
these ambiguities can be listed
as transformations ($\Theta_1 \ldots \Theta_{12}$) whereby 
a simultaneous modification of the coefficients as indicated below has 
no physical meaning:
\ba
 \Theta_1: && 
 \delta \omega^{ }_1 = - \delta \psi^{ }_1 = \delta \psi^{ }_{10} \;,
 \la{Theta_1} \\ 
 \Theta_2: && 
 \delta \omega^{ }_4 = - \delta \psi^{ }_4 = \delta \psi^{ }_{13} \;, \\ 
 \Theta_3: && 
 \delta \omega^{ }_5 = - \delta \psi^{ }_5 = \delta \psi^{ }_{16} \;, \\ 
 \Theta_4: && 
 \delta \omega^{ }_{22} = - \delta \psi^{ }_{22} = \delta \psi^{ }_{30} \;, \\ 
 \Theta_{5}: && 
 \delta \omega^{ }_{23} = - \delta \omega^{ }_{24} =
 \delta \omega^{ }_{31} = - \delta \psi^{ }_{23} = \delta \psi^{ }_{24}
 =  2 \delta \psi^{ }_{39} = -2 \delta \psi^{ }_{40} \;, \hspace*{6mm} \\ 
 \Theta_6: && 
 \delta \omega^{ }_{25} = - \delta \psi^{ }_{25} = \delta \psi^{ }_{38} \;, \\ 
 \Theta_7: && 
 \delta \omega^{ }_{26} = - \delta \psi^{ }_{26} = \delta \psi^{ }_{34} \;, \\ 
 \Theta_{8}: && 
 \delta \omega^{ }_{27} = - \delta \omega^{ }_{28} =
 \delta \omega^{ }_{35} = - \delta \psi^{ }_{27} = \delta \psi^{ }_{28}
 = 2 \delta \psi^{ }_{43} = -2 \delta \psi^{ }_{44} \;, \hspace*{6mm} \\ 
 \Theta_9: && 
 \delta \omega^{ }_{29} = - \delta \psi^{ }_{29} = \delta \psi^{ }_{42} \;, \\ 
 \Theta_{10}: && 
 \delta \psi^{ }_{13} = \delta \psi^{ }_{15} =
  - \delta \psi^{ }_{16} = - \delta \psi^{ }_{18} \;, \\ 
 \Theta_{11}: && 
 \delta \psi^{ }_{30} = \delta \psi^{ }_{31} =
  - \delta \psi^{ }_{38} = - 2\delta \psi^{ }_{39}
 = - 2\delta \psi^{ }_{40} \;, \\ 
 \Theta_{12}: && 
 \delta \psi^{ }_{34} = \delta \psi^{ }_{35} =
  - \delta \psi^{ }_{42} = - 2\delta \psi^{ }_{43} = - 2\delta \psi^{ }_{44} 
 \;. \la{Theta_12}
\ea
This list may not be complete. It can be checked that the expressions
in \eqs\nr{MQCD_1l} and 
\nr{C1}--\nr{C3} are invariant in these transformations. 

The 5-point Chapman vertex reads
\ba
 \delta S^{(5)}_\rmii{EQCD}  & = & 
  A^a_\mu(q)\, A^b_\nu(r)\, A^c_\rho(s)\, A^d_\alpha(t)\, A^e_\beta(u)
  \, \delta(q+r+s+t+u)  
 \biggl(  \Tint{P}'\frac{i \gE^5 s^{ }_\mu }{P^6} \biggr)
 \nn & \times & \biggl\{ 
 X^{\{ab\}[cde]}\, 
 \biggl[
  \kappa^{ }_1 \, S^{ }_{\rho\alpha} S^{ }_{\nu\beta}
 + \kappa^{ }_2 \, S^{ }_{\rho\beta} S^{ }_{\nu\alpha}
 + \kappa^{ }_3 \, S^{ }_{\rho\nu} S^{ }_{\alpha\beta}
 \nn & & \hspace*{18mm}
 +\, \kappa^{ }_4 \, T^{ }_{\rho\alpha} S^{ }_{\nu\beta}
 + \kappa^{ }_5 \, T^{ }_{\rho\beta} S^{ }_{\nu\alpha}
 + \kappa^{ }_6 \, T^{ }_{\rho\nu} S^{ }_{\alpha\beta}
 \nn & & \hspace*{18mm}
 +\, \kappa^{ }_7 \, S^{ }_{\rho\alpha} T^{ }_{\nu\beta}
 + \kappa^{ }_8 \, S^{ }_{\rho\beta} T^{ }_{\nu\alpha}
 + \kappa^{ }_9 \, S^{ }_{\rho\nu} T^{ }_{\alpha\beta}
 + \kappa^{ }_{10} \, T^{ }_{\rho\nu \alpha\beta} \biggl] 
 \nn & + & 
 X^{[ab]\{cde\}}\, 
 \bigl[ \kappa^{ }_i \to \lambda^{ }_i \bigr]
 \biggr\}
 \;, \la{5pt_ops}
\ea
where
\ba
 && \kappa^{ }_1 = -8 c^{ }_1 \;, \quad
    \kappa^{ }_2 = 32 c^{ }_1 \;, \quad
    \kappa^{ }_3 = -8 c^{ }_1 \;, \nn
 && \kappa^{ }_4 = -8 c^{ }_1 -8 c^{ }_2\;, \quad
    \kappa^{ }_5 = 32 c^{ }_1 +32 c^{ }_2\;, \quad
    \kappa^{ }_6 = -8 c^{ }_1 -8 c^{ }_2\;, \nn
 && \kappa^{ }_7 = -8 c^{ }_1 -8 c^{ }_2 \;, \quad
    \kappa^{ }_8 = 32 c^{ }_1 +8 c^{ }_5 - 16 c^{ }_7\;, \quad
    \kappa^{ }_9 = -8 c^{ }_1 -8 c^{ }_2 \;, \nn 
 &&   \kappa^{ }_{10} = 16 c^{ }_1  + 8 c^{ }_5-16 c^{ }_7 - 8 c^{ }_9\;,  \nn
 && \lambda^{ }_1 = 40 c^{ }_1 - 24 c^{ }_3 \;, \quad
    \lambda^{ }_2 = -32 c^{ }_1 + 24 c^{ }_3\;, \quad
    \lambda^{ }_3 = 24 c^{ }_1 \;, \nn
 && \lambda^{ }_4 = 40 c^{ }_1  + 8 c^{ }_2
  - 24 c^{ }_3-8c^{ }_4 + 8 c^{ }_5 \;, \quad
    \lambda^{ }_5 = -32 c^{ }_1 + 24 c^{ }_3 + 8 c^{ }_4 - 8 c^{ }_5\;, \nn
 &&  \lambda^{ }_6 = 24 c^{ }_1 + 24 c^{ }_2  \;, \quad 
    \lambda^{ }_7 = 40 c^{ }_1  + 8 c^{ }_2
  - 24 c^{ }_3-8c^{ }_4 + 8 c^{ }_5 \;, \nn
 &&  \lambda^{ }_8 = -32 c^{ }_1 -32 c^{ }_2
  + 24 c^{ }_3 + 8 c^{ }_4 - 16 c^{ }_7\;, \quad
    \lambda^{ }_9 = 24 c^{ }_1 - 8 c^{ }_2 + 8 c^{ }_5 + 16 c^{ }_6 \;, \nn 
 &&   \lambda^{ }_{10} = 32 c^{ }_1 + 16 c^{ }_5
  + 16 c^{ }_6 -16 c^{ }_7 + 8 c^{ }_8 - 8 c^{ }_9
 \;. \la{sc_5}
\ea
Finally the 6-point vertex can be expressed as 
\ba
 \delta S^{(6)}_\rmii{EQCD}  & = & 
 \int_X 
  A^a_\mu\, A^b_\nu\, A^c_\rho\, A^d_\sigma\,
  A^e_\alpha\, A^f_\beta
  \, X^{abcde\!f}
 \biggl(  \Tint{P}'\frac{\gE^6 }{P^6} \biggr)
 \nn & \times & \Bigl\{ 
 \bigl[
  \chi^{ }_1 \, S^{ }_{\rho\sigma} S^{ }_{\alpha\beta}
 + \chi^{ }_2 \, S^{ }_{\rho\alpha} S^{ }_{\sigma\beta}
 + \chi^{ }_3 \, S^{ }_{\rho\beta} S^{ }_{\sigma\alpha}
 \bigr] \, S^{ }_{\mu\nu}
 \nn & & \hspace*{0mm}
 + \bigl[
   \chi^{ }_4 \, S^{ }_{\nu\alpha} S^{ }_{\rho\beta}
 + \chi^{ }_5 \, S^{ }_{\nu\beta} S^{ }_{\rho\alpha}
 \bigr] \, S^{ }_{\mu\sigma}
 \nn & & \hspace*{0mm}
 + \bigl[
  \chi^{ }_6 \, S^{ }_{\rho\sigma} S^{ }_{\alpha\beta}
 + \chi^{ }_7 \, S^{ }_{\rho\alpha} S^{ }_{\sigma\beta}
 + \chi^{ }_8 \, S^{ }_{\rho\beta} S^{ }_{\sigma\alpha}
 \bigr] \, T^{ }_{\mu\nu}
 \nn & & \hspace*{0mm}
 + \bigl[
   \chi^{ }_9 \, S^{ }_{\nu\sigma} S^{ }_{\alpha\beta}
 + \chi^{ }_{10} \, S^{ }_{\nu\alpha} S^{ }_{\sigma\beta}
 \bigr] \, T^{ }_{\mu\rho}
 \nn & & \hspace*{0mm}
 + \bigl[
   \chi^{ }_{11} \, S^{ }_{\nu\rho} S^{ }_{\alpha\beta}
 + \chi^{ }_{12} \, S^{ }_{\nu\alpha} S^{ }_{\rho\beta}
 + \chi^{ }_{13} \, S^{ }_{\nu\beta} S^{ }_{\rho\alpha}
 \bigr] \, T^{ }_{\mu\sigma}
 \nn & & \hspace*{0mm}
 +  \chi^{ }_{14} \, S^{ }_{\mu\nu} T^{ }_{\rho\sigma\alpha\beta}
 + \chi^{ }_{15} \, S^{ }_{\mu\rho} T^{ }_{\nu\sigma\alpha\beta}
 + \chi^{ }_{16} \, S^{ }_{\mu\sigma} T^{ }_{\nu\rho\alpha\beta} 
 + \chi^{ }_{17} \, T^{ }_{\mu\nu\rho\sigma\alpha\beta}
 \Bigr\}
 \;, \la{6pt_ops}
\ea
where
\ba
 && \chi^{ }_1 = -4 c^{ }_1 + 2 c^{ }_3 \;, \quad
    \chi^{ }_2 = 16 c^{ }_1 - 6 c^{ }_3 \;, \quad
    \chi^{ }_3 = -4 c^{ }_1 \;, \quad
    \chi^{ }_4 = -2 c^{ }_3\;, \quad
    \chi^{ }_5 = -8 c^{ }_1 +6 c^{ }_3\;, \nn
 && \chi^{ }_6 = -12 c^{ }_1 -4 c^{ }_2
  +6 c^{ }_3 + 2c^{ }_4 -2 c^{ }_5\;, \quad
    \chi^{ }_7 = 16 c^{ }_1 -6 c^{ }_3-2c^{ }_4 + 4 c^{ }_5 + 4c^{ }_6 \;, \nn
 &&   \chi^{ }_8 = -8 c^{ }_1 -2 c^{ }_5 - 4 c^{ }_6\;, \nn 
 &&  \chi^{ }_{9} = 32 c^{ }_1  + 16 c^{ }_2
  - 12 c^{ }_3 -4 c^{ }_4 + 4 c^{ }_5\;,  \quad
   \chi^{ }_{10} = -16 c^{ }_1  + 12 c^{ }_3 + 4 c^{ }_4 - 4 c^{ }_5\;,  \nn
 &&  \chi^{ }_{11} = -4 c^{ }_1  -4 c^{ }_2 \;,  \quad
   \chi^{ }_{12} = -6 c^{ }_3 -2 c^{ }_4 + 4 c^{ }_7\;,  \quad
   \chi^{ }_{13} = -8 c^{ }_1  -8 c^{ }_2
 + 6 c^{ }_3 + 2 c^{ }_4 - 4 c^{ }_7\;,  \nn
 &&  \chi^{ }_{14} = -4 c^{ }_1  - 2 c^{ }_5 - 4 c^{ }_6 - 2 c^{ }_8\;,  \quad
   \chi^{ }_{15} = 16 c^{ }_1  + 8 c^{ }_5
  + 8 c^{ }_6 - 8 c^{ }_7 + 4 c^{ }_8 - 4 c^{ }_9\;,  \nn
 &&  \chi^{ }_{16} = -12 c^{ }_1  -6 c^{ }_5 -4 c^{ }_6
  + 8 c^{ }_7 - 2 c^{ }_8 + 4 c^{ }_9 \;, \quad
   \chi^{ }_{17} = - 2 c^{ }_{10} 
 \;. \la{sc_6}
\ea

%
\section{Basic vacuum integrals}

For the computations of \se\ref{se:scalepiT} various $d$-dimensional
vacuum integrals are needed. At 2-loop level 
their results can be expressed in 
terms of $H$ defined in \eq\nr{H}, multiplied by rational
functions of $d$. 
For notational simplicity we denote the 
mass by $m$, let 
$
 \Delta_p \equiv p^2 + m^2
$, and omit the trivial factor $T$ included in \eq\nr{H}.

Making use of the integral 
\be
 \int_{p} \frac{1}{\Delta_p^n} = 
 \frac{m^{d - 2n} \Gamma(n - \frac{d}{2})}
      {(4\pi)^{\frac{d}{2}} \Gamma(n)}
 \;, \la{masterI}
\ee
factorized integrals can be expressed as 
\ba
 && 
 \int_{{p},{q}} \frac{ m^{-2} }{\Delta_p \Delta_q} = 
 - \frac{2(d-3)H}{d-2}
 \;, \quad
 \int_{{p},{q}} \frac{1}{\Delta_p^2 \Delta_q} = 
 (d-3)H
 \;.
\ea
A sunset integral with a power of the massless propagator reads
\be
 \int_{{p},{q}} \frac{1}{\Delta_p \Delta_q ({p+q})^{2n} } = 
 \frac{m^{2d - 2n -4} \Gamma(\frac{d}{2}-n) \Gamma(n+2-d) 
      \Gamma^{2}(n+1-\frac{d}{2})}
      {(4\pi)^{{d}} \Gamma(\frac{d}{2}) \Gamma(2n+2 - d)}
 \;.
\ee
In particular, 
\be
  \int_{{p},{q}} \frac{1}{\Delta_p \Delta_q ({p+q})^2}
 = H
 \;, \quad
  \int_{{p},{q}} \frac{m^2}{\Delta_p \Delta_q ({p+q})^4}
 = -\frac{(d-3)H}{2(d-5)}
 \;. \la{H_var}
\ee
A sunset integral with a power of a massive propagator reads
\be
 \int_{{p},{q}} \frac{1}{\Delta_p^n \Delta_q ({p+q})^2} = 
 \frac{m^{2d - 2n -4} \Gamma(1-\frac{d}{2}) \Gamma(n+1-\frac{d}{2})}
      {(d-n-2) (4\pi)^{{d}} \Gamma(n)}
 \;.
\ee
In particular, 
\be
  \int_{{p},{q}} \frac{m^2}{\Delta^2_p \Delta_q ({p+q})^2}
 = -\frac{(d-3)H}{2}
 \;, \quad
  \int_{{p},{q}} \frac{m^4}{\Delta^3_p \Delta_q ({p+q})^2}
 = \frac{(d-3)(d-4)(d-6)H}{8(d-5)}
 \;. 
\ee

Tensor integrals can be reduced to scalar integrals through 
\ba
 && 
 \langle p^{ }_\mu p^{ }_\nu p^{ }_\alpha p^{ }_\beta \rangle
 = \frac{
    ( S^{ }_{\mu\nu}S^{ }_{\alpha\beta} + 
     S^{ }_{\mu\alpha}S^{ }_{\nu\beta} + 
     S^{ }_{\mu\beta}S^{ }_{\nu\alpha} ) \langle p^4 \rangle}{d(d+2)}
 \;, \\ 
 && 
 \langle p^{ }_\mu p^{ }_\nu p^{ }_\alpha q^{ }_\beta \rangle
 = \frac{
    ( S^{ }_{\mu\nu}S^{ }_{\alpha\beta} + 
     S^{ }_{\mu\alpha}S^{ }_{\nu\beta} + 
     S^{ }_{\mu\beta}S^{ }_{\nu\alpha} )
    \langle p^2 {p}\cdot{q} \rangle}{d(d+2)}
 \;, \\
 && 
 \langle p^{ }_\mu p^{ }_\nu q^{ }_\alpha q^{ }_\beta \rangle
 = \frac{
    (S^{ }_{\mu\alpha}S^{ }_{\nu\beta} + 
     S^{ }_{\mu\beta}S^{ }_{\nu\alpha} )
    \langle d({p}\cdot{q})^2 - p^2 q^2 \rangle}{d(d-1)(d+2)}
 + 
 \frac{
     S^{ }_{\mu\nu}S^{ }_{\alpha\beta}
    \langle  (d+1) p^2 q^2 -2 ( {p}\cdot{q})^2  \rangle}{d(d-1)(d+2)}
 \;, \nn 
\ea
where $\langle ... \rangle_{ }$ represents a generic rotationally
invariant expectation value, and 
$S^{ }_{\mu\nu} \equiv \delta^{ }_{\mu i} \delta^{ }_{\nu i}$. 

In the considerations of \se\ref{ss:MQCD}, another variant
of the sunset integral was encountered, 
\be
 H^{ }_3 \; \equiv \; 
 \int_{{p},{q}} \frac{1}{\Delta_p \Delta_q \Delta_{p+q}}
 \;. \la{H3} 
\ee
It can be written
in terms of the hypergeometric 
function ${}_2F_1$~\cite{Davydychev:1992mt,Schroder:2005va},
\be
H^{ }_3 
  =  -\frac{3(d-2)}{4(d-3)}\biggl[
 {}_2F_1\Big(\frac{4-d}2,1;\frac{5-d}2;\frac34\Big)
 -3^{\frac{d-5}2}\frac{2\pi\Gamma(5-d)}{\Gamma(\frac{4-d}2)\Gamma(\frac{6-d}2
 )}\biggr]
 \int_{{p},{q}} \frac{ m^{-2} }{\Delta_p \Delta_q}
 \;.
\ee

At 3-loop level we need the values of two ``basketball'' 
integrals~(cf.\ e.g.\ refs.~\cite{aminusb,akr}): 
\ba
 B^{ }_2 & \equiv & 
 \int_{p,q,r} \frac{1}{\Delta^{ }_p \Delta^{ }_q (p+r)^2(q+r)^2}
 \nn 
 & = & 
 -\frac{m\mu^{-6\epsilon}}{(4\pi)^3}
 \biggl( \frac{\bmu}{2m} \biggr)^{6\epsilon}_{ }
 \biggl\{ 
   \frac{1}{2\epsilon} + 4 + \epsilon \, 
   \biggl[ 26 + \frac{25\zeta^{ }_2}{4} \biggr] + \rmO(\epsilon^2)
 \biggr\} 
 \;, \la{b_2} \\ 
 B^{ }_4 & \equiv & 
 \int_{p,q,r} \frac{1}{\Delta^{ }_p \Delta^{ }_q
  \Delta^{ }_{p+r} \Delta^{ }_{q+r} }
 \nn 
 & = & 
 -\frac{m\mu^{-6\epsilon}}{(4\pi)^3}
 \biggl( \frac{\bmu}{2m} \biggr)^{6\epsilon}_{ }
 \biggl\{ 
   \frac{1}{\epsilon} + 8 - 4\ln 2 + \epsilon \, 
   \biggl[ 52 + \frac{17\zeta^{ }_2}{2} 
 - 32 \ln 2 + 4 \ln^2 2 \biggr] + \rmO(\epsilon^2)
 \biggr\} 
 \;. \nn \la{b_4}
\ea

%
\section{Details concerning 2-loop and 3-loop results}

For completeness we report here technical results related to 
\ses\ref{se:scalepiT} and \ref{se:scalemE} that were too lengthy
to fit the presentation in the main text. 

Consider first the coefficients $C^{ }_1, C^{ }_2$ and $C^{ }_3$, 
defined in \eq\nr{2loopgM5}.
Because of the general way in which we have parametrized 
the Chapman vertices (cf.~appendix~C), the expressions for these 
contain substantial ``redundancies'', which we reproduce here in full. 
This permits for very strong 
crosschecks, as discussed e.g.\ in the context 
of \eqs\nr{Theta_1}--\nr{Theta_12} for the quartic Chapman vertex. 
The expressions read
\ba
 C^{ }_1 & = & 
  - \frac{8(d-1)\bigl[ 
  (2d+3)\varzeta^{ }_1
 +2d(d+2) \varzeta^{ }_2
 +(d+1)\xi^{ }_5 
 - (d+2) \xi^{ }_6 
 - \xi^{ }_7 
 + d\, \xi^{ }_{10} 
 \bigr]}{d-2}
 \nn  & - &
 \frac{8(d-1)\bigl[ 
 (d+1)(d+2) \xi^{ }_8 - (d^2 + 3d + 1)\xi^{ }_9
 \bigr]}{d-2}
 \nn & + & 
 \frac{2(d-1)\bigl[ 
  4(\psi^{ }_3 - \psi^{ }_{30} + \psi^{ }_{31})
  -2 (2d+3) \psi^{ }_{10} + 4 d \psi^{ }_{12} 
  - 3 \psi^{ }_{22} + \omega^{ }_{22} 
  \bigr]}{d-2}
 \nn & - & 
 \frac{(d-1)\bigl[ 
    2(3d^2 - 1) \psi^{ }_4
 + 4  (2d^2+1) (\psi^{ }_{13} - \psi^{ }_{15})
  + (5d-1) \psi^{ }_{26}
  + d (\psi^{ }_{27} -\omega^{ }_{27} ) 
  \bigr]}{d-2}
 \nn & - & 
 \frac{(d-1)\bigl[ 
 \psi^{ }_6 - \omega^{ }_6
  + \psi^{ }_{28} - \omega^{ }_{28}  
 + 2 (5d+1) (\psi^{ }_{34} - \psi^{ }_{35})
 - 2 (d^2 + 3 ) \omega^{ }_4 - (5d + 3 ) \omega^{ }_{26} 
  \bigr]}{d-2}
 \nn  & - &
 \frac{(d-1)\bigl[
  (3 d + 7) (\kappa^{ }_4 + 2 \psi^{ }_1)
  + (d - 1) ( 2 \kappa^{ }_5 - \lambda^{ }_4 - 2 \omega^{ }_1 + 
              2 \omega^{ }_{35} )
  - 5 \kappa^{ }_6 - (4 d + 1) \lambda^{ }_6
  \bigr]}{d-2}
 \nn & - & 
 \frac{10d(d-3)\bigl[ 
   \kappa^{ }_{10} - \lambda^{ }_{10 }
  - 4 \chi^{ }_{14} - 2\chi^{ }_{15} - 2\chi^{ }_{16}
  + 4\psi^{ }_{19} - 2 \psi^{ }_{21}
 \bigr]}{d-2}
 \;, \la{C1} \\[3mm] 
 C^{ }_2 & = & 
 \frac{2\bigl[ 
 18 (d-1) \xi^{ }_4
 +(d+1)(d^2 - 9 d + 12) ( \xi^{ }_6 - \xi^{ }_5 )
 + 12 (d^2 - 3) \xi^{ }_7
 \bigr]}{3(d-5)}
 \nn  & - &
 \frac{2\bigl( 
   d^6 - 13 d^5 + 49 d^4 - 83 d^3 + 208 d^2 - 114 d - 156
 \bigr) \varzeta^{ }_2}{3(d-5)(d-7)}
 \nn  & - &
 \frac{\bigl(
   4 d^5 - 55 d^4 + 226 d^3 - 335 d^2 + 484 d - 336
 \bigr)\xi^{ }_8}{3(d-5)(d-7)} 
 \nn  & + &
 \frac{\bigl( 
   4 d^5 - 55 d^4 + 226 d^3 - 323 d^2 + 388 d - 252
 \bigr) \xi^{ }_9}{3(d-5)(d-7)}  
 \nn  & - &
 \frac{4 \bigl( 
  d^4 - 10 d^3 + 25 d^2 - 51 d + 51
 \bigr) \varzeta^{ }_1}{3(d-5)}
  - 
 \frac{2 \bigl(
  2 d^4 - 31 d^3 + 120 d^2 - 111 d + 36
 \bigr) \xi^{ }_{10}}{3(d-5)}
 \nn & + & 
 \frac{(d-1)\bigl[ 
  (3d+7) \psi^{ }_1 -4 (\psi^{ }_3 - \psi^{ }_{30} + \psi^{ }_{31})
  + 2 (2d+3) \psi^{ }_{10} -4 d \psi^{ }_{12} + 3 \psi^{ }_{22} 
 \bigr]}{d-5}
 \nn & + & 
 \frac{(d-1)\bigl[ 
  \psi^{ }_{28}
  - 2 (d-1) \omega^{ }_1 -2 \omega^{ }_{22} - \omega^{ }_{28} 
 \bigr]}{2(d-5)}
  + \frac{d(37d-39) \psi^{ }_5}{6}
  - \frac{d(3d-1)\omega^{ }_5}{2}
 \nn 
 & + & \frac{(d-2)(d-3)(d-7)
  (\psi^{ }_4 + 3 \omega^{ }_4 - 2 \psi^{ }_{13})}{6(d-5)}
 - \frac{(d^3 - 8 d^2 + 51 d-84)\psi^{ }_6}{12(d-5)}
 \nn & + &
 \frac{2(d^2 - 8d + 9 )\psi^{ }_{15} }{3(d-5)}
 +  \frac{d(23d-21)\psi^{ }_{16}}{3}
  - 2  (4d^2 - 5d + 2) \psi^{ }_{18} 
 \nn & + & 
 \frac{(d^2 + 7d - 12) 
 (\psi^{ }_{26} - 2 \psi^{ }_{34} + 3 \omega^{ }_{26})}{12}
 + \frac{d(d+1)\psi^{ }_{35}}{6}
  -  2(d-2) \psi^{ }_{44}
 \nn & - & 
 \frac{(d^3 - 16 d^2 + 59 d - 52)\omega^{ }_6}{4(d-5)}
  -  \frac{(d-2)\bigl[ 
    (d^2-33) \psi^{ }_{27}
  - (d^2 - 24 d + 87) \omega^{ }_{27}
 \bigr]}{12(d-5)}
 \nn  & + &
 \frac{d(d-3)\bigl[ 
  5 ( \lambda^{ }_{10} - \kappa^{ }_{10} ) 
  -20 (d-2) \psi^{ }_{19} + 4 (2d-3) \psi^{ }_{21} - \omega^{ }_{35} 
 \bigr]}{6}
 \nn & + &
 \frac{\alpha (d-1)\bigl[ \psi^{ }_{28} - \omega^{ }_{28} - 
 2 \omega^{ }_{35} - 8 (2 \varzeta^{ }_1
 + \xi^{ }_5 + \xi^{ }_7 )\bigr]}{2(d-5)} 
 \nn & + &
  \frac{4 \alpha(d-1) \xi^{ }_8}{d-7}
 + \frac{8 \alpha(d-1)\bigl[(d-3)\varzeta^{ }_2
 -  \xi^{ }_9 \bigr]}{(d-5)(d-7)}
 \;, \la{C2} \\[3mm] 
 C^{ }_3 & = & 
 8d(d-1)\bigl[ \varzeta^{ }_2 + \xi^{ }_8 + \xi^{ }_{10} \bigr]
 \nn  & + &
 \frac{4 (d-1) \bigl[ 
 (d-1)(\varzeta^{ }_1 + \xi^{ }_5)
 +   2( \xi^{ }_2 + \xi^{ }_3 + \xi^{ }_4 + \xi^{ }_6 ) 
 + (d+1) \xi^{ }_7
 \bigr]}{d-5}
 \nn  & + &
 \frac{(d-1)\bigl[ 
  (3d+7)(\psi^{ }_1 + \psi^{ }_{25})
  - 4 (\psi^{ }_3 + \psi^{ }_{23} + \psi^{ }_{24} + \psi^{ }_{31})
  + 2 (2d+3)(\psi^{ }_{10} + \psi^{ }_{38})
 \bigr]}{d-5} 
 \nn  & - &
 \frac{(d-1)\bigl[ 
   4 d (\psi^{ }_{12} + \psi^{ }_{39} + \psi^{ }_{40})
  - 10 ( \psi^{ }_{22} + \psi^{ }_{30}) 
  + (d-1)(\omega^{ }_{1} + \omega^{ }_{25}) 
 \bigr]}{d-5} 
 \nn & + & 
 2d(d-1)
 \bigl[
   3(\psi^{ }_5 + \psi^{ }_{29})
  + 4(\psi^{ }_{16} - \psi^{ }_{18} + \psi^{ }_{42} - \psi^{ }_{43}
  - \psi^{ }_{44})
  - \omega^{ }_5 - \omega^{ }_{29}  
 \bigr]
 \;. \la{C3}
\ea
After substituting the coefficients from appendix~C,  
we get \eq\nr{2loopgM5_2}. 

As a second ingredient, we report the full $d$-dimensional version
of \eq\nr{gM2_3l_bare}. The result can be expressed as 
\ba
 \delta^{ } \widetilde{\Gamma}^{(2)}_\rmii{MQCD} [B] & = &
 \fr12 B^a_i (q)\, B^b_j (r)  \, \delta^{ab} \, \delta(q+r) \, 
 \bigl(  q^2 \delta^{ }_{ij} - q^{ }_i q^{ }_j  \bigr)
 \biggl( \frac{\gE^2 \Nc^{ } }{\mE^{2}} \biggr)^3
  \la{fulld_first} \\ 
 & \times & 
 \Bigl\{
    \bigl( r^{ }_1 + \tilde{r}^{ }_1  \bigr) (d)\, I^3(\mE^{ }) 
 + r^{ }_2(d)\, \mE^2 B^{ }_2(\mE^{ })
 +  \bigl( r^{ }_3 + \tilde{r}^{ }_3 \bigr) (d) \, 
      \mE^2 B^{ }_4(\mE^{ })
 \Bigr\}
 \;, \nonumber
\ea 
where the pure gauge contributions are parametrized by 
\ba
  r^{ }_1(d)&=&
  -\frac{(d-2)p^{ }_1(d)}
   {384(d-10)(d-8)(d-7)(d-6)(d-5)(d-4)(d-3)^2(d-1)d}
 \;, \\ 
  r^{ }_2(d)&=&
   \frac{(3d-10)(3d-8)p^{ }_2(d)}{128(d-3)(d-1)d(2d-11)(2d-9)(2d-7)}
 \;, \\ 
  r^{ }_3(d)&=&
  \frac{(3d-10)(3d-8)p^{ }_3(d)}{256(d-10)(d-8)(d-6)(d-4)(d-1)d}
 \;, 
\ea
with the non-factorizable polynomials
\ba
  p^{ }_1(d)&=&
  12d^{\,12}-628d^{\,11}+14447d^{\,10}
 -193505d^{\,9}+1689420d^{\,8}-10234582d^{\,7}
 \nn &+&
    44883931d^{\,6}-147059385d^{\,5}+366585830d^{\,4}-689809244d^{\,3}
 \nn &+&
    929595256d^{\,2}-791686464d+314842752 
 \;, \\ 
  p^{ }_2(d)&=&
  12d^{\,7}-308d^{\,6}+3175d^{\,5}-17441d^{\,4}+57347d^{\,3}
 \nn &-& 117419d^{\,2}+138786d-70872
 \;, \\ 
  p^{ }_3(d)&=&
  3d^{\,5}-60d^{\,4}+359d^{\,3}-670d^{\,2}+400d+736
 \;, 
\ea
where $I, B^{ }_2$ and $B^{ }_4$
are the master integrals from 
\eqs\nr{I}, \nr{b_2} and \nr{b_4}, respectively. 
In terms of the couplings from \eqs\nr{sc_a}--\nr{sc_c}, 
the scalar contributions amount to 
\ba
  \tilde{r}^{ }_1(d)&=&
  \frac{d-2}{8}\biggl\{ 
  \frac{(d-4)(3d^{\,5}-49d^{\,4}+283d^{\,3}-779d^{\,2}+1238d-1056)\lambda}
  {3(d-7)(d-5)(d-3)d}
 \nn & & \hspace*{1cm} 
  -\,\frac{(d-4)(3d-10)\lambda^2}{3}
 \nn & & \hspace*{1cm} 
  +\,\frac{(d-2)^2(9d^{\,2}-77d+158)\kappa^{ }_1}{16(d-6)(d-4)(d-3)d}
  +\frac{(d-10)(d-2)^2 \kappa^{ }_2}{16(d-4)d}
  \biggr\} 
  \;, \\[2mm] 
  \tilde{r}^{ }_3(d)&=&
  \frac{(3d-10)(3d-8)(d^{\,2}-5d-2)
   \bigl[ \kappa^{ }_1+(d-6)\kappa^{ }_2\bigr]
  }{256(d-6)(d-4)d}
  \;. \la{fulld_last}
\ea
Setting $d=3-2\epsilon$, inserting the values of the master integrals,
and carrying out a Taylor expansion
in $\epsilon$, \eq\nr{fulld_first} goes over into \eq\nr{gM2_3l_bare}. 

\small{
%

}

\end{document}